\documentclass[letterpaper]{article} 
\usepackage[]{aaai2027}  
\usepackage[hyphens]{url}  
\usepackage{graphicx} 
\usepackage{natbib}  
\usepackage{caption} 
\usepackage{algorithm}
\usepackage{algorithmic}

\usepackage{newfloat}
\usepackage{listings}
\DeclareCaptionStyle{ruled}{labelfont=normalfont,labelsep=colon,strut=off} 
\floatstyle{ruled}
\newfloat{listing}{tb}{lst}{}
\floatname{listing}{Listing}

\usepackage{booktabs}

\title{Causal Explanations for Stratified Datalog}
\author{
    Written by AAAI Press Staff\textsuperscript{\rm 1}\thanks{With help from the AAAI Publications Committee.}\\
    AAAI Style Contributions by Peter Patel Schneider,
    Sunil Issar,\\
    J. Scott Penberthy,
    George Ferguson,
    Hans Guesgen,
    Francisco Cruz\equalcontrib\corresponding,
    Marc Pujol-Gonzalez\equalcontrib\corresponding
}
\affiliations{
    \textsuperscript{\rm 1}Association for the Advancement of Artificial Intelligence\\

    1101 Pennsylvania Ave, NW Suite 300\\
    Washington, DC 20004 USA\\
    proceedings-questions@aaai.org
}

\title{Causal Explanations for Stratified Datalog \thanks{With Extended Proofs and Supplementary Results}}
\author{
    Ratan Bahadur Thapa\textsuperscript{\rm 1} \thanks{Corresponding Author: ratan.thapa@ki.uni-stuttgart.de},\quad
    Steffen Staab\textsuperscript{\rm 1,\rm 2}
}
\affiliations {
    \textsuperscript{\rm 1}Institute for Artificial Intelligence, University of Stuttgart,  Germany\\
    \textsuperscript{\rm 2}University of Southampton, United Kingdom 
}
\usepackage[hyphens]{url}
\usepackage{natbib}
\usepackage{amsmath,amssymb,mathtools}
\usepackage{amsthm}
\usepackage{cleveref}
\usepackage{xcolor}
\usepackage{accsupp}
\newtheorem{definition}{Definition}
\newtheorem{theorem}{Theorem}
\newtheorem{proposition}{Proposition}

\newtheorem{corollary}{Corollary}
\newtheorem{example}{Example}

\newcommand\mf[1]{\ensuremath{\mathfrak{#1}}}
\newcommand\tx[1]{\texttt{#1}}

\newcommand\ma[1]{\ensuremath{\mathcal{#1}}}
\newcommand\mb[1]{\ensuremath{\mathbb{#1}}}

\newcommand{\PiP}{\Pi}
\newcommand{\Dx}{D^{\mathsf{x}}}
\newcommand{\Univ}{\mathcal{U}}
\newcommand{\EDB}{\Sigma_{\mathsf{E}}}
\newcommand{\IDB}{\Sigma_{\mathsf{I}}}
\newcommand{\PM}{\ensuremath{\mf{P}}}
\newcommand{\Ans}{\textsf{ans}}
\newcommand{\Prime}{\textsf{Prime}}
\newcommand{\Supp}{\textsf{Supp}}
\newcommand{\Trans}{\tx{Tr}}
\newcommand{\Dual}{\textsf{Dual}}
\newcommand{\Rob}{\textsf{Rob}}
\newcommand{\Resp}{\textsf{Resp}}
\newcommand{\Kappa}{\kappa}
\newcommand{\Goal}{\mathsf{Goal}}

\newcommand{\Path}{\tx{Path}}
\newcommand{\Edge}{\tx{Edge}}
\newcommand{\Block}{\tx{Block}}
\newcommand{\toggle}{\mathbin{\triangle}}
\newcommand{\Dist}{\tx{d}}
\newcommand{\Inf}{\infty}
\newcommand{\Cause}{\textsc{Cause}}
\newcommand{\Robust}{\textsc{Robust}}
\newcommand{\Responsibility}{\textsc{Responsibility}}
\newcommand{\VC}{\textsc{Vertex Cover}}

\newcommand{\Min}{\tx{Min}}
\newcommand{\merge}{\mathbin{\sqcup}}
\newcommand{\primeor}{\mathbin{\vee_{\tx{PI}}}}
\newcommand{\ind}{\mathbf{1}}
\newcommand{\PowTwo}[1]{\BeginAccSupp{method=escape,ActualText={2\string^#1}}2^{#1}\EndAccSupp{}}

\newcommand{\Rel}{\tx{Rel}}
\newcommand{\Implicant}{\textsc{Implicant}}
\newcommand{\ResponseEquivalence}{\textsc{Response-Equivalence}}
\newcommand{\Flip}{\textsf{Flip}}
\newcommand{\Circ}{\mb C}

\newcommand{\OppDist}{\delta^{\tx{opp}}}

\newcommand{\PiEX}{\Pi_{\mathsf{ex}}}

\begin{document}
\nocopyright
\maketitle

\begin{abstract}
Rule-based reasoning with exceptions requires causal explanations that account for both present and absent facts. For positive Datalog, monotonicity allows minimal supports to determine deletion causality. Stratified negation removes that property because inserting or deleting a fact may create or destroy an answer. We study
actual causes, responsibility, and robustness for safe  and stratified Datalog
under perfect-model semantics and interventions over a finite set of mutable
extensional facts. We prove that minimal supports and inclusion-minimal
outcome-changing interventions do not determine causality, while robustness
radius one may coexist with unbounded minimum contingencies. Our main result characterizes the minimum contingency size of a candidate fact by compatible prime implicants for the observed and opposite outcomes. The characterization conservatively recovers support-based causality for positive Datalog and yields path--cut characterizations for blocked recursive reachability. For fixed nonrecursive stratified programs, we establish data-complexity
NP-completeness for cause recognition, robustness, and responsibility, and
coNP-completeness for intervention-response equivalence between two such programs.
\end{abstract}

\section{Introduction}\label{sec:introduction}
Rule-based systems often derive conclusions subject to exceptions.  An
eligibility decision may require a qualification and the absence of an
exclusion; an access decision may hold unless a revocation is derived; and a
diagnosis may remain valid unless a contraindication is established. Explaining such conclusions requires reasoning about both present and absent facts.
Stratified datalog gives such rules a deterministic declarative semantics while
retaining recursion and polynomial-time data evaluation
\cite{abiteboul1995foundations,apt1988towards}.  We study causal explanation for
a ground conclusion over a finite set of mutable extensional facts.  At an
observed input, we ask which present or absent fact can become pivotal under a
contingency, how strongly it is responsible, and how many changes are required
to reverse the answer.  

To make these questions concrete, we begin with an example that introduces the relevant causal quantities. Consider the Datalog program $\Pi_{\mathsf{ex}}$
\[
\begin{array}{rcl}
 \mathsf{approve}(x)&\leftarrow&
 \mathsf{eligible}(x),\neg\mathsf{blocked}(x),\\
 \mathsf{blocked}(x)&\leftarrow&\mathsf{violation}(x),\\
 \mathsf{blocked}(x)&\leftarrow&
 \mathsf{highRisk}(x),\neg\mathsf{reviewed}(x).
\end{array}
\tag{$\star$}\label{eq:example}
\]
Let $D^x=\varnothing$,
$E=\{\mathsf{highRisk}(a)\}$,
$A=\mathsf{approve}(a)$, and
\[
\Univ=\{\mathsf{eligible}(a),\mathsf{violation}(a),
       \mathsf{highRisk}(a),
        \mathsf{reviewed}(a)\},
\]
where $D^x$ is the fixed extensional database whose facts cannot be
changed, $E$ is the observed mutable input, $\Univ$ is the finite set of
mutable extensional facts that may be inserted or deleted, and $A$ is the
ground goal. An intervention $J\subseteq\Univ$ is applied by changing the membership in $E$
of every atom in $J$. Thus, the intervention toggles every atom in $J$,
inserting those absent from $E$ and deleting those present in $E$.

At $E$, the goal $A$ is false. The singleton intervention
$J=\{\mathsf{eligible}(a)\}$ leaves it false: since
$\mathsf{highRisk}(a)$ is present and $\mathsf{reviewed}(a)$ is absent, the
last rule derives $\mathsf{blocked}(a)$. Inserting
$\mathsf{reviewed}(a)$ alone also leaves $A$ false because eligibility is
absent. However, inserting both $\mathsf{reviewed}(a)$ and
$\mathsf{eligible}(a)$ makes $A$ true, because eligibility and review are
present and violation is absent. Thus, $\mathsf{eligible}(a)$ is an actual
cause of the false outcome with a contingency
$\Gamma=\{\mathsf{reviewed}(a)\}$ for that cause. The responsibility
of a cause with minimum contingency size $k$ is $1/(k+1)$. Since the
contingency $\Gamma$ is minimum and $|\Gamma|=1$, the responsibility of
$\mathsf{eligible}(a)$ is $1/(|\Gamma|+1)=1/2$. Robustness is the minimum
size of an intervention that reverses the observed outcome. Therefore, the
robustness value is $2$ in this instance because the above intervention of
size $|J|=2$ makes $A$ true, whereas no singleton intervention does.

The program $\Pi_{\mathsf{ex}}$ also explains why absence conditions matter.
The absence of $\mathsf{eligible}(a)$ at $E$ is sufficient for the false
outcome. The true outcome requires $\mathsf{eligible}(a)$ and the absence of
$\mathsf{violation}(a)$, together with either the absence of
$\mathsf{highRisk}(a)$ or the presence of $\mathsf{reviewed}(a)$.
The false and true conditions disagree on the value of
$\mathsf{eligible}(a)$. Once this atom is treated as the candidate cause,
either deleting $\mathsf{highRisk}(a)$ or inserting
$\mathsf{reviewed}(a)$ before toggling it yields a minimum contingency
of size one. To illustrate
deletion-based causality, consider instead
\[
E'=\{\mathsf{eligible}(a),\mathsf{highRisk}(a),
     \mathsf{reviewed}(a)\}.
\]
Then, $\mathsf{approve}(a)$ holds under $\Pi_{\mathsf{ex}}$ at the
alternative observed state $E'$. A fact in the observed
state is a deletion cause if deleting it reverses the outcome, possibly after
a contingency that alone preserves the outcome. Deleting
$\mathsf{reviewed}(a)$ activates the blocking rule and therefore makes the
approval false. Thus, in the alternative observed state $E'$,
$\mathsf{reviewed}(a)$ is a deletion cause with the empty contingency
$\Gamma=\varnothing$.

This nonmonotonic behavior shows why the positive-Datalog support representation
does not extend to stratified negation. A minimal support is an
inclusion-minimal set of mutable facts deriving an answer; by monotonicity, an
answer holds exactly when its input contains one. In the deletion-based causality, the resulting support hypergraph determines deletion causes,
responsibility, and robustness
\cite{thapa2026causalityminimalsupportsrecursive}. Under stratified
negation, minimal supports omit necessary absences, and
inclusion-minimal outcome-changing interventions can omit
atoms pivotal only at states farther from the observation. We prove that neither representation determines actual
causes or responsibilities and that a robustness radius of one can coexist
with unbounded minimum contingencies.

Database causality defines actual causes, contingencies, and responsibility for
query answers and non-answers
\cite{meliou2010causality,meliou2010complexity}, while prime implicants encode
inclusion-minimal partial assignments forcing a Boolean outcome
\cite{darwiche2022computation,dubslaff2022causality}. Feature-causality accounts
minimize simultaneous changes containing the candidate without requiring the
remaining changes alone to preserve the observed outcome. We instead pair
compatible observed- and opposite-outcome implicants containing complementary
candidate literals. Compatibility enforces this intermediate-state condition,
and minimum pair cost gives the minimum contingency size for every finite
Boolean response. For stratified Datalog, we construct both prime families
compositionally through stratum-wise perfect-model evaluation. For blocked
recursive reachability, we derive path--cut pairs and an exact responsibility
formula.


Our contributions are as follows: (i) in
Section~\ref{sec:representation}, we prove that minimal supports and
inclusion-minimal outcome-changing interventions do not determine causality,
establish an unbounded separation between robustness and minimum contingency
size, and prove the compatible prime-pair characterization and its positive
Datalog specialization; (ii) in Section~\ref{sec:reachability}, we derive the
recursive path--cut characterization and exact responsibility formula, and show
that explicit prime families may be exponential although the corresponding
causal quantities remain polynomial-time computable; and (iii) in
Section~\ref{sec:complexity},  we establish data-complexity bounds for fixed programs and goal predicates, with the database-level objects supplied as input.

\section{Related Work}\label{sec:related}

Database causality studies causes, contingencies, and responsibility for answers
and non-answers
\cite{meliou2010causality,meliou2010complexity,bertossi2017causes}. For
conjunctive queries, answer causality typically deletes present tuples, whereas
non-answer causality inserts missing tuples. We allow both operations at one
observed state in recursive stratified programs. Closely related work uses minimal supports and transversals to characterize
deletion-based causality, responsibility, and robustness for positive recursive
Datalog \cite{thapa2026causalityminimalsupportsrecursive}.
Corollary~\ref{cor:positive} recovers those positive-
Datalog characterizations from compatible prime implicants. Resilience and deletion
propagation also study minimum deletions for monotone queries
\cite{freire2015complexity,buneman2002propagation}.

Responsibility measures for conjunctive queries with negation include weighted
sums of minimal supports \cite{bienvenu2026responsibility}. We instead study
pivotality under arbitrary contingencies and mixed interventions, including
recursion. Theorem~\ref{thm:abstraction-separation} shows that minimal supports
do not determine the minimum contingency size of a designated atom.

Provenance games and graphs represent successful and failed rule
applications under negation
\cite{kohler2013first,lee2017efficientlycomputingprovenancegraphs}.
Semiring approaches represent recursive derivations and input annotations
\cite{deutch2014circuits,bourgaux2022revisiting,
gradel2025provenance,zhao2024evaluating}. Prime implicants instead capture minimal partial assignments
forcing the semantic outcome.

Prime implicants also characterize sufficient reasons and feature explanations
\cite{darwiche2022computation,dubslaff2022causality}. Existing feature accounts
minimize sets of features changed simultaneously, including the candidate. Our
contingency semantics additionally requires that changing only the noncandidate
atoms in the set preserves the observed outcome before the candidate is toggled.
Compatible prime pairs enforce this intermediate-state requirement. Unlike structural-model and
causal probabilistic approaches
\cite{halpern2005causes,chockler2004responsibility,vennekens2011actual}, we treat
the program as a deterministic specification, intervene on extensional atoms,
and evaluate the goal in its perfect model.

\section{Datalog and Interventions}\label{sec:datalog}

Let $\EDB$ and $\IDB$ be disjoint finite signatures of extensional and
intensional predicates, respectively, each with fixed arity. We allow nullary predicates and use
nullary $\IDB$ predicates as Boolean goals. A safe Datalog rule with negation has the form
\[
H\leftarrow B_1,\ldots,B_m,\neg C_1,\ldots,\neg C_n,
\]
where $H$ is the $\IDB$ head atom, the $B_i$ are positive body atoms, the
$\neg C_j$ are negative body literals, each $B_i$ and $C_j$ is an
$\EDB$- or $\IDB$-atom, and every variable in $H$ and $C_j$ occurs
in some $B_i$. A finite Datalog program $\PiP$ is stratified if some map
$\lambda:\EDB\cup\IDB\to\mathbb N$, with $\lambda(S)=0$ for every
$S\in\EDB$, satisfies $\lambda(S)\leq\lambda(R)$ whenever a rule with
head predicate $R$ contains a positive body atom with predicate $S$, and
$\lambda(S)<\lambda(R)$ whenever it contains a negative body literal
with predicate $S$. Stratum-wise least-fixpoint evaluation yields the
unique perfect model $\PM(\PiP\cup D)$ for every finite $\EDB$-database
$D$ \cite{apt1988towards,abiteboul1995foundations}. We call $\PiP$ safe
and stratified if every rule in $\PiP$ is safe and $\PiP$ is stratified.
\begin{definition}\label{def:instance}
An intervention instance is a tuple
\[
 I=(\PiP,\Dx,\Univ,E,A),
\]
where $\PiP$ is a safe stratified program, $\Dx$ is a finite exogenous
$\EDB$-database, $\Univ$ is a finite set of mutable ground
$\EDB$-atoms disjoint from $\Dx$, $E\subseteq\Univ$ is the distinguished observed
state, and $A$ is a ground $\IDB$-atom.  Any $F\subseteq\Univ$ is a state and represents exactly the mutable atoms
present as facts, and we
write
\[
 \Ans_I(F)=
 \begin{cases}
 1,&A\in\PM(\PiP\cup\Dx\cup F),\\
 0,&A\notin\PM(\PiP\cup\Dx\cup F).
 \end{cases}
 \tag{1}\label{eq:response}
\]
\end{definition}

The active domain contains the constants in $\PiP,\Dx,\Univ$, and $A$.
We omit the subscript $I$ when the instance is understood.  For an intervention $J\subseteq\Univ$ and a state $F\subseteq\Univ$, applying $J$ yields
$F\toggle J=(F\setminus J)\cup(J\setminus F)$.  Each atom in $J$ is toggled: atoms in $F\cap J$ are deleted, and atoms in $J\setminus F$ are inserted.  The finite domain $\Univ$ specifies the admissible interventions.

\begin{example}\label{ex:states}
Consider the program $\PiEX$ in~\eqref{eq:example}.  Let
\[
\begin{aligned}
\Univ &=  \{\mathsf{eligible}(a),\mathsf{violation}(a),
    \mathsf{highRisk}(a),\mathsf{reviewed}(a)\},\\
E &=\{\mathsf{highRisk}(a)\} \,\,\text{and}\,\, A =\mathsf{approve}(a),
\end{aligned}
\]
and let $I_{\mathsf{ex}} = (\PiEX,\varnothing,\Univ,E,A)$.
Since $\mathsf{eligible}(a)\notin E$, we have
$\Ans_{I_{\mathsf{ex}}}(E)=0$. Let
$J=\{\mathsf{reviewed}(a)\}$.
Applying $J$ to the observed state gives
\[F= E\toggle J = \{\mathsf{highRisk}(a),\mathsf{reviewed}(a)\}.
\]
Since $\mathsf{eligible}(a)\notin F$, the answer remains false:
$\Ans_{I_{\mathsf{ex}}}(F)=0$.
Now let $J'=\{\mathsf{eligible}(a)\}$.
Applying $J'$ to $F$ gives 
\[F\toggle J'
=
\{\mathsf{eligible}(a),\mathsf{highRisk}(a),
  \mathsf{reviewed}(a)\}.
\]
In this state, $\mathsf{eligible}(a)$ and $\mathsf{reviewed}(a)$ hold,
whereas $\mathsf{violation}(a)$ does not hold.  Thus, $\Ans_{I_{\mathsf{ex}}}(F\toggle J')=1$. Hence, the additional intervention $J'$ changes
the answer from false to true.
\end{example}

\begin{definition}\label{def:cause}
Let  $b=\Ans(E)$ and $\tau\in\Univ$, where $b\in\{0,1\}$ is the observed outcome at $E$.  The mutable atom $\tau$ is an actual
cause of outcome $b$ at $E$ if some contingency
$\Gamma\subseteq\Univ\setminus\{\tau\}$ satisfies
\[\Ans(E\toggle\Gamma)=b
 \,\,\,\text{and}\,\,\,
 \Ans(E\toggle(\Gamma\cup\{\tau\}))=1-b.
 \tag{2}\label{eq:cause}\]
It is counterfactual if $\Gamma=\emptyset$ satisfies
\eqref{eq:cause}.  Let $\Kappa_E(\tau)$ be the minimum contingency size,
with value $\Inf$ when none exists, and let the responsibility of $\tau$ at $E$
\[
 \Resp_E(\tau)=
 \begin{cases}
 1/(\Kappa_E(\tau)+1),&\Kappa_E(\tau)<\Inf,\\
 0,&\Kappa_E(\tau)=\Inf.
 \end{cases}
 \tag{3}\label{eq:responsibility}
\]
The robustness radius is
\[
 \Rob(E)=\min\{\,|J|\,\,\mid\Ans(E\toggle J)=1-b\,\},
 \tag{4}\label{eq:robustness-definition}
\]
with value $\Inf$ when $\Ans$ is constant.
\end{definition}

Every quantity is relative to $\Univ$: enlarging the mutable domain may
introduce new contingencies.  We therefore compare programs only over a common
intervention domain.

\section{Causal Representation}\label{sec:representation}
We first prove that support-based summaries lose contingency information under
negation.  We then state the Boolean representation needed for the exact
causal characterization.

For $S\subseteq\Univ$, we call $S$ a positive support if $\Ans(S)=1$,
and minimal if no proper subset has outcome $1$.  Let $\Supp_I$ be the
family of minimal supports.  For the observed state $E$, let
\[
 \Flip_I(E)=
 \Min_{\subseteq}\{J\subseteq\Univ\, \mid \Ans_I(E\toggle J)\neq\Ans_I(E)\},
\]
where $\Min_{\subseteq}$ selects the inclusion-minimal members.

\begin{theorem}\label{thm:abstraction-separation}
There are intervention instances $I_1$ and $I_2$ with fixed
nonrecursive stratified programs and common $(\Dx,\Univ,E,A)$ such that
$\Supp_{I_1}=\Supp_{I_2}$ and
$\Flip_{I_1}(E)=\Flip_{I_2}(E)$, but their actual-cause families and
responsibilities at $E$ differ.
\end{theorem}
\begin{example}\label{example:two_programs}
Let $p=P(c)$, $q=Q(c)$, and $r=R(c)$. Consider
$\Dx=\emptyset$, $\Univ=\{p,q,r\}$, $E=\emptyset$, $A=\Goal$ and the response functions $f_1,f_2 : 2^\Univ \to \{0,1\}$
\[
\begin{aligned}
 f_1(F)=1
 &\Longleftrightarrow
 p\in F\land(q\in F\Longleftrightarrow r\in F),\\
 f_2(F)=1
 &\Longleftrightarrow p\in F.
\end{aligned}
\tag{5}\label{eq:two-responses-main}
\]
Then, the following program $\PiP_1$  computes the first response:
\[
\begin{array}{rcl}
 \Goal&\leftarrow&P(c),\neg Q(c),\neg R(c),\\
 \Goal&\leftarrow&P(c),Q(c),R(c).
\end{array}
\tag{6}\label{eq:separation-program}
\]
Let $\PiP_2=\{\Goal\leftarrow P(c)\}$ and
$I_j=(\PiP_j,\Dx,\Univ,E,A)$ for $j\in\{1,2\}$.  The program $\PiP_2$
computes $f_2(F)$.  Both have the unique minimal
support $\{p\}$ and
$\Flip_I(E)=\{\{p\}\}$.  In the $\PiP_1$, however,
$q$ and $r$ are causes with contingencies $\{p,r\}$ and
$\{p,q\}$, respectively, and each has responsibility $1/3$.  The response $f_2$ is independent of $q$ and $r$.  Thus, supports and all
inclusion-minimal interventions do not determine contingency causality.   
\end{example}

The separation persists at unbounded contingency distance.

\begin{theorem}\label{thm:remote-contingency}
There is a fixed nonrecursive stratified program such that, for every
$n\geq2$, an intervention instance of size $O(n)$ has observed outcome
$0$, robustness radius $1$, one inclusion-minimal outcome-changing
intervention, and $n$ further actual causes, each with minimum contingency
size $n$ and responsibility $1/(n+1)$.
\end{theorem}
\begin{example}
Let $a_1,\ldots,a_n$ be exogenous items, let $g$ be a mutable gate, and
let $s_i=\mathsf{Sel}(a_i)$ be mutable selections, and let
$S=\{s_1,\ldots,s_n\}$.  The fixed program is
\[
\begin{array}{rcl}
 \mathsf{Some}&\leftarrow&\mathsf{Item}(x),\mathsf{Sel}(x),\\
 \mathsf{Missing}&\leftarrow&\mathsf{Item}(x),\neg\mathsf{Sel}(x),\\
 \mathsf{None}&\leftarrow&\neg\mathsf{Some},
 \qquad\mathsf{All}\leftarrow\neg\mathsf{Missing},\\
 \Goal&\leftarrow&g,\mathsf{None},
 \qquad\Goal\leftarrow g,\mathsf{All}.
\end{array}
\tag{7}\label{eq:remote-program}
\]
For every mutable state $F$, perfect-model evaluation gives
\[
\begin{aligned}
 \Ans(F)=1\,\Longleftrightarrow\,
&g\in F\, \wedge\,
 F\cap S\in\{\emptyset,S\}.
\end{aligned}
\]
At $E=\emptyset$, toggling $g$ alone derives the goal through
$\mathsf{None}$, so $\Rob(E)=1$ and $\{g\}$ is the unique
inclusion-minimal outcome-changing intervention.  For $s_i$, the contingency
$\{g\}\cup\{s_j\mid j\neq i\}$ produces a mixed selection and preserves
outcome $0$; toggling $s_i$ makes every item selected and derives the
goal through $\mathsf{All}$.  Any pivotal state for $s_i$ must contain
the gate and every other selection, proving minimum size $n$.  Thus, a
unit robustness radius and a unique inclusion-minimal outcome-changing intervention provide no bound
on the contingency size of another cause; the corresponding nonzero
responsibility tends to zero.  In particular, for every $\varepsilon>0$
there is such an instance with $\Rob(E)=1$, one inclusion-minimal outcome-changing intervention,
and a cause of responsibility below $\varepsilon$. 
\end{example}

A term over $\Univ$ is a pair $C=(P,N)$ of disjoint subsets of
$\Univ$.  It represents
$\bigwedge_{\tau\in P}\tau\wedge\bigwedge_{\tau\in N}\neg\tau$: a member
$\tau\in P$ represents the positive Boolean literal $\tau$, and a member $\tau\in N$
represents the negative Boolean literal $\neg\tau$. A state $F$ satisfies $C$, written
$F\models C$, if $P\subseteq F$ and $N\cap F=\emptyset$.  We write
$(P',N')\preceq(P,N)$ if $P'\subseteq P$ and $N'\subseteq N$.
For $\tau\in\Univ$, let $\ell_E(\tau)=\tau$ if $\tau\in E$ and
$\ell_E(\tau)=\neg\tau$ otherwise; let
$\ell_E^{\mathsf c}(\tau)$ denote its complementary literal.

\begin{definition}\label{def:prime}
For $b\in\{0,1\}$, a term $C$ is a $b$-implicant if
\[
 \forall F\subseteq\Univ,\,\, F\models C\,\Longrightarrow\,\Ans(F)=b.
\]
It is prime if no proper subterm is a $b$-implicant.  The family of all
prime $b$-implicants is denoted by $\Prime_I^b$.
\end{definition}
\noindent We also call prime $1$-implicants prime truth conditions and prime $0$-implicants prime falsity conditions.

For the rule
$\Goal\leftarrow P(c),\neg Q(c)$, the unique prime truth condition is
$(\{P(c)\},\{Q(c)\})$, while the prime falsity conditions are
$(\emptyset,\{P(c)\})$ and $(\{Q(c)\},\emptyset)$.  Thus, negative and positive literals necessary for truth and
falsity, respectively.

Two terms conflict if one contains a literal and the other contains its complement; otherwise, they are compatible. For a term family $\mathcal C$, let $\Dual(\mathcal C)$ contain
the $\preceq$-minimal terms that conflict with every member of
$\mathcal C$. The following finite Boolean fact is standard
\cite{crama2011boolean}.

\begin{proposition}\label{prop:representation}
For every state $F\subseteq\Univ$ and $b\in\{0,1\}$,
\[
 \Ans(F)=b
 \,\,\Longleftrightarrow\,\,
 \exists C\in\Prime_I^b\;(F\models C)
 \tag{8}\label{eq:prime-cover},
\]
\[
 \Prime_I^{1-b}=\Dual(\Prime_I^b).
 \tag{9}\label{eq:prime-duality}
\]
\end{proposition}

Theorem~\ref{thm:remote-contingency} has an exact prime representation.
Its two prime truth conditions are
$C_{\mathsf{none}}=(\{g\},S)$ and $
 C_{\mathsf{all}}=(\{g\}\cup S,\emptyset),$
and its prime falsity conditions are
\[
 (\emptyset,\{g\})
 \quad\text{and}\quad
 (\{s_i\},\{s_j\})
 \quad(i\neq j).
\]
The first falsity term closes the gate; every other falsity term forces a mixed
selection.  At $E=\emptyset$, choose $i\neq j$.  The literal
$\ell_E(s_i)=\neg s_i$ occurs in $(\{s_j\},\{s_i\})$, whereas
$C_{\mathsf{all}}$ contains its complementary literal $s_i$.  Replacing $s_i$
by $\neg s_i$ in $C_{\mathsf{all}}$ and merging the terms requires $g$ and all
$s_j$ with $j\neq i$.  Their distance from $E$ is $n$, recovering the
minimum contingency in Theorem~\ref{thm:remote-contingency}.  Thus, the
opposite-prime pairing captures a cause that neither minimal truth conditions
at the observed state nor nearest outcome changes retain.


Prime families admit a finite compositional iteration, although the
explicit term families may be exponential.
For a ground intensional atom $H$, let $\Prime_I^b(H)$ denote the prime
$b$-implicants of its Boolean response. 
Let $m_i$ be the number of
ground intensional atoms in stratum $i$.

\begin{proposition}\label{prop:compositional}
For every stratum $i$ and every ground intensional atom $H$ in that
stratum, the simultaneous prime-family iteration reaches
$(\Prime_I^1(H),\Prime_I^0(H))$ after at most $m_i$ stages.
\end{proposition}

The proposition is a supporting construction rather than the causal
characterization: it establishes that the semantic prime families can be
obtained from stratified evaluation, while Theorem~\ref{thm:causal-characterization}
identifies the information needed for minimum contingencies.

For brevity, let $e=\mathsf{eligible}(a)$, $v=\mathsf{violation}(a)$, $h=\mathsf{highRisk}(a)$, and $r=\mathsf{reviewed}(a)$. For $I_{\mathsf{ex}}$ from Example~\ref{ex:states}, direct minimization gives
\[
\begin{aligned}
 \Prime_{I_{\mathsf{ex}}}^1(\mathsf{approve}(a))
 &=\bigl\{(\{e\},\{v,h\}),(\{e,r\},\{v\})\bigr\},\\
 \Prime_{I_{\mathsf{ex}}}^0(\mathsf{approve}(a))
 &=\bigl\{(\emptyset,\{e\}),(\{v\},\emptyset),(\{h\},\{r\})\bigr\}.
\end{aligned}
\tag{10}\label{eq:approval-primes}
\]
The terms describe semantic input conditions rather than particular rule
derivations.

For a term $C=(P,N)$, let
\[
 \Dist_E(C)=|P\setminus E|+|N\cap E|.
 \tag{11}\label{eq:term-distance}
\]
The value is the number of literals of $C$ not satisfied by $E$, equivalently
the minimum number of toggles required to reach a state satisfying $C$.  Let
$C^\tau$ replace the literal on $\tau$ by its complement when $C$ contains a
literal on $\tau$.

For compatible terms, let
$(P,N)\merge(P',N')=(P\cup P',N\cup N')$.  We write
\[
 \Dist_{E,-\tau}(P,N)=
 |(P\setminus E)\setminus\{\tau\}|+
 |(N\cap E)\setminus\{\tau\}|.
\]
The minimum of an empty family is $\Inf$. Let $\mathcal Q_{E,\tau}^b$ contain all pairs
$(C,D)\in\Prime_I^b\times\Prime_I^{1-b}$ such that $C$ contains
$\ell_E(\tau)$, $D$ contains $\ell_E^{\mathsf c}(\tau)$, and
$C$ is compatible with $D^\tau$.  Let
\[
 K_E(\tau)=
 \min_{(C,D)\in\mathcal Q_{E,\tau}^b}
 \Dist_{E,-\tau}(C\merge D^\tau),
 \tag{12}\label{eq:pair-distance}
\]

\begin{theorem}\label{thm:causal-characterization}
Let $b=\Ans(E)$ and $\tau\in\Univ$.
\begin{enumerate}
\item The mutable atom $\tau$ is an actual cause at $E$ iff some
$C\in\Prime_I^b$ contains $\ell_E(\tau)$.
\item $\Kappa_E(\tau)=K_E(\tau)$. The responsibility is
$\Resp_E(\tau)=1/(K_E(\tau)+1)$ when $K_E(\tau)<\Inf$, and
$\Resp_E(\tau)=0$ otherwise.
\item
\[\Rob(E)=\min_{D\in\Prime_I^{1-b}}\Dist_E(D).
 \tag{13}\label{eq:robustness-prime}
\]
\end{enumerate}
\end{theorem}

Item~(2) is the central representation result.  For the upper bound,
take $C,D$ from $\mathcal Q_{E,\tau}^b$ and toggle exactly the
noncandidate atoms whose literals in $C\merge D^\tau$ are not satisfied by $E$.
The resulting state satisfies $C$, retains outcome $b$, and its
$\tau$-neighbour satisfies $D$, with outcome $1-b$.  For the
lower bound, let $\Gamma$ be a minimum contingency. By
Equation~\eqref{eq:prime-cover}, choose a prime $b$-implicant $C$ satisfied by
$E\toggle\Gamma$ and a prime $(1-b)$-implicant $D$ satisfied by the adjacent
state.  The terms contain complementary literals on $\tau$, are compatible after
replacing the literal on $\tau$ in $D$ by its complement, and contain at most
$|\Gamma|$ literals outside $\tau$ that are not satisfied by $E$.  Both inequalities yield
$K_E(\tau)=\Kappa_E(\tau)$.  The cause and robustness clauses use the
same adjacent-state construction.  

The compatible current-outcome term in Equation~\eqref{eq:pair-distance} is
essential.  Let
\[\begin{aligned}
 \OppDist_E(\tau)=& \min\{\Dist_{E,-\tau}(D) \mid D\in\Prime_I^{1-b},\\
 & D\text{ contains }\ell_E^{\mathsf c}(\tau)\}.
\end{aligned}
 \tag{14}\label{eq:opposite-only}\]
The quantity considers only a nearest opposite-outcome condition.

\begin{proposition}\label{prop:compatibility-necessity}
There is a fixed nonrecursive stratified program and an intervention instance
with an actual cause $\tau$ such that
$\OppDist_E(\tau)=1$ but $\Kappa_E(\tau)=2$.
\end{proposition}

\begin{example}
Let $\Univ=\{p,q,r\}$, $E=\{p,r\}$ (abbreviations as in Eq.~\ref{eq:two-responses-main}), and let $\PiP_3$ be the fixed program containing
\[
\begingroup
\begin{array}{rcl}
 \Goal&\leftarrow&\neg P(c),\neg R(c),\\[-0.5ex]
 \Goal&\leftarrow&\neg Q(c),\neg R(c),
\end{array}
\tag{15}\label{eq:compatibility-example}
\endgroup
\]
Let $I_3=(\PiP_3,\emptyset,\Univ,E,\Goal)$.  Its response is $\neg r\wedge(\neg p\vee\neg q)$.  The prime falsity
conditions are $(\{r\},\emptyset)$ and $(\{p,q\},\emptyset)$, and the
prime truth conditions are $(\emptyset,\{p,r\})$ and
$(\emptyset,\{q,r\})$.  For candidate atom $p$, the nearest opposite
condition $(\emptyset,\{p,r\})$ contains outside $p$ only the literal
$\neg r$, which is not satisfied by $E$; thus
$\OppDist_E(p)=1$.  Toggling only $r$, however, reaches $\{p\}$,
where the outcome is already true.  The contingency $\{q,r\}$ reaches
$\{p,q\}$, where the outcome remains false, and toggling $p$ reaches
$\{q\}$, where it becomes true.  Neither singleton contingency works;
therefore, $\Kappa_E(p)=2$.  The compatible pair
$(\{p,q\},\emptyset)$ and $(\emptyset,\{p,r\})$ recovers the two required
contingency changes.  Thus, nearest opposite conditions alone do not determine
responsibility.

For $\Pi_1$ from Example~\ref{example:two_programs}, at $E=\emptyset$,
pairing the prime $0$-implicant $(\{q\},\{r\})$, which contains
$\ell_E(r)=\neg r$, with the prime $1$-implicant
$(\{p,q,r\},\emptyset)$ yields the compatible term
$(\{p,q\},\{r\})$. Equation~\eqref{eq:pair-distance} then gives
$\Kappa_E(r)=2$ and $\Resp_E(r)=1/3$, although $\{p\}$ is the only
inclusion-minimal outcome-changing intervention.
\end{example}
The preceding examples derive causal quantities from compatible prime implicants, however, these quantities do not determine the full intervention response at one observed state.  We write $\ind[\varphi]$ for the indicator of
$\varphi$.  Let $\Univ=\{p,q,r\}$ and $E=\{p\}$.  The responses
\begin{enumerate}
    \item $f_1(F)=\ind[p\in F]$ and
    \item $f_2(F)=\ind[p\in F\text{ or }\{q,r\}\subseteq F]$
\end{enumerate}
have the same sole cause $p$, with responsibility $1$, and robustness
radius $1$ at $E$, but disagree at $F=\{q,r\}$. 

Let 
$I=(\PiP,\Dx,\Univ,E,A) $ and $
I'=(\PiP',\Dx,\Univ,E,A)$
be instances that differ only in their programs.  We call the replacement
of $\PiP$ by $\PiP'$ intervention-preserving if
$\Ans_I(F)=\Ans_{I'}(F)$ for every $F\subseteq\Univ$.
\begin{corollary}\label{cor:rewrite}
The replacement is intervention-preserving iff
$\Prime_I^1=\Prime_{I'}^1$, and equivalently iff
$\Prime_I^0=\Prime_{I'}^0$.
\end{corollary}

 Corollary~\ref{cor:positive} recovers the positive-Datalog support
characterization \cite{thapa2026causalityminimalsupportsrecursive}. For a family $\mathcal S\subseteq 2^{\Univ}$, let
\[ \Trans(\mathcal S)=
 \Min_{\subseteq}\{T\subseteq\Univ\mid\forall S\in\mathcal S,\ T\cap S\neq\emptyset\}.\]
We use $\Trans(\emptyset)=\{\emptyset\}$ and
$\Trans(\mathcal S)=\emptyset$ when $\emptyset\in\mathcal S$.

\begin{corollary}\label{cor:positive}
If $\PiP$ is positive, then
\[\begin{aligned}
 \Prime_I^1&=\{(S,\emptyset) \mid S\in\Supp_I\},\\
 \Prime_I^0&=\{(\emptyset,T) \mid T\in\Trans(\Supp_I)\}.
\end{aligned}
\tag{16}\label{eq:positive-specialization}\]
At $E=\Univ$, Theorem~\ref{thm:causal-characterization}
therefore gives the positive-Datalog characterizations of causes,
responsibility, and deletion robustness.
\end{corollary}


\section{Blocked Reachability}\label{sec:reachability}
We now derive a recursion-specific theory in which the two prime families have
distinct graph meanings.

Consider the fixed stratified program $\Pi_{\mathsf{br}}$:
\[\noindent\begin{aligned}
 \Path(x,y)&\leftarrow \Edge(x,y),\neg\Block(x,y), \\
 \Path(x,y)&\leftarrow \Edge(x,z),\neg\Block(x,z),\Path(z,y).
\end{aligned}\tag{17}\label{eq:blocked-path}
\]
Let $H=(V,L)$ be a finite simple directed graph of potential edges.
Let $s,t\in V$ be distinct.  A potential path is a nonempty simple
directed $s$-$t$ path in $H$, written $\pi:s\leadsto t$.  An edge
cut is a set $K\subseteq L$ meeting every potential path; $K=\emptyset$
is permitted.  For $e=(u,v)\in L$, write $E_e=\Edge(u,v)$ and
$B_e=\Block(u,v)$, and let $\Univ_H=\{E_e,B_e\mid e\in L\}$.
For $E\subseteq\Univ_H$, consider
$I_{H,E}=(\Pi_{\mathsf{br}},\emptyset,\Univ_H,E,\Path(s,t))$.
Let
$L_F=\{e\in L\mid E_e\in F\text{ and }B_e\notin F\}$.  An edge is active in
state $F$ exactly when it belongs to $L_F$, and $H_F=(V,L_F)$ is the
active graph.
Presence of $E_e$ and absence of $B_e$ enable $e$; absence of
$E_e$ and presence of $B_e$ disable $e$.  For a potential path
$\pi$, let
\[
 C_\pi=(\{E_e \mid e\in\pi\},\{B_e \mid e\in\pi\}).
\]
For an inclusion-minimal directed $s$-$t$ edge cut $K$, let
$\alpha:K\to\{\mathsf{edge},\mathsf{block}\}$ select one disabling
literal for each cut edge. We call the pair $(K,\alpha)$ a \emph{mode-labelled cut}, and let
\[\begin{aligned}
 D_{(K,\alpha)}=(&\{B_e \mid e\in K,\ \alpha(e)=\mathsf{block}\},\\
               &\{E_e \mid e\in K,\ \alpha(e)=\mathsf{edge}\}).
\end{aligned}\]
For a state $E$ whose active graph contains a directed $s$-$t$
path, let $\lambda_{H_E}(s,t)$ denote the minimum size of a directed
$s$-$t$ edge cut in $H_E$.

\begin{theorem}\label{thm:path-cut}
For every finite simple directed graph $H=(V,L)$, every pair of distinct
vertices $s,t\in V$, and every state $E\subseteq\Univ_H$, the associated
blocked-reachability instance $I_{H,E}$ satisfies:
\begin{enumerate}
\item the prime $1$-implicants are exactly the terms $C_\pi$;
\item the prime $0$-implicants are exactly the terms $D_{(K,\alpha)}$;
\item if $H_E$ contains an $s$-$t$ path, then
\[
 \Rob(E)=\lambda_{H_E}(s,t),
 \tag{18}\label{eq:cut-radius}
\]
\item if $H_E$ contains no $s$-$t$ path, then
\[
 \Rob(E)=
 \min_{\pi:s\leadsto t}
 \sum_{e\in\pi}
 \bigl(\ind[E_e\notin E]+\ind[B_e\in E]\bigr),
 \tag{19}\label{eq:path-radius}
\]
where the minimum of an empty path family is $\Inf$;
\item if $\Ans(E)=1$, then $E_e$ is a cause exactly when
$E_e\in E$ and $e$ lies on a potential $s$-$t$ path, and
$B_e$ is a cause exactly when $B_e\notin E$ and $e$ lies on such a
path; if $\Ans(E)=0$, then $E_e$ is a cause exactly when
$E_e\notin E$ and $e$ lies on a potential path, and $B_e$ is a
cause exactly when $B_e\in E$ and $e$ lies on a potential path. 
\end{enumerate}
\end{theorem}

For the truth direction, every $C_\pi$ forces $\pi$ to be active.
Conversely, let $C=(P,N)$ be a prime truth implicant and complete it by
making every unmentioned edge absent and every unmentioned blocker present.
The completion satisfies $C$, so it contains an active path $\pi$.
Every enabling literal of $\pi$ already occurs in $C$; hence,
$C_\pi\preceq C$, and primeness gives $C=C_\pi$.  For falsity,
complete a prime term by making every unmentioned edge present and every
unmentioned blocker absent.  Its explicitly disabled edges meet every
directed $s$-$t$ path and therefore contain a cut.  Primeness retains an
inclusion-minimal cut and exactly one disabling literal per cut edge, yielding
$D_{(K,\alpha)}$.

The nearest opposite term gives the robustness formulas. Destroying a true
answer disables a- set of active edges forming an $s$--$t$ cut in $H_E$,
whereas creating a false answer activates a potential path.  The cause clauses select the literals $\ell_E(\tau)$ occurring in
the corresponding path or cut terms. 

Consider one directed diamond with paths
$\pi_1=(x_1,x_2)$ and $\pi_2=(y_1,y_2)$.  Its prime truth
conditions are $C_{\pi_1}$ and $C_{\pi_2}$.  The inclusion-minimal edge cuts are
$K_{ij}=\{x_i,y_j\}$ such that $i,j\in\{1,2\}$. Each of the four cuts admits four mode labellings.  For example,
\[D_{(K_{11},\alpha)} =(\{B_{x_1}\},\{E_{y_1}\})
\]
after choosing blocker presence on $x_1$ and edge absence on $y_1$.
The term forces both paths to fail while leaving every other edge and blocker
unconstrained.  Removing either literal reactivates a path in a suitable
completion, thus the term is prime.  

The two robustness clauses differ even on small graphs.  Suppose
$\lambda_{H_E}(s,t)=2$.  Toggling the two edge facts of a minimum cut
changes the answer, while no single toggle does; therefore,
$\Rob(E)=2$.  For a false state with one potential path
$\pi=(f_1,f_2)$, assume $E_{f_1}\in E$, $B_{f_1}\notin E$,
$E_{f_2}\notin E$, and $B_{f_2}\in E$.  The path has activation
cost two: inserting $E_{f_2}$ and deleting $B_{f_2}$ creates the
answer.  Equation~\eqref{eq:path-radius} proves minimality without
enumerating interventions.  Blockers participate symmetrically.  If a unique
potential path has all edge facts present and exactly one blocker $B_e$
present, the answer is false and deleting $B_e$ activates the path.  Thus,
$B_e$ is counterfactual for the non-answer.  If the blocker is absent in a
true state, inserting it can instead destroy the answer after a contingency
disables all alternative paths.

For a path $\pi$, let
\[ a_E(\pi)=\sum_{f\in\pi}
 \bigl(\ind[E_f\notin E]+\ind[B_f\in E]\bigr)\]
be its activation cost.  For a mode-labelled cut $(K,\alpha)$, let
\[c_E(K,\alpha)=
 \sum_{f\in K}
 \begin{cases}
  \ind[E_f\in E],&\alpha(f)=\mathsf{edge},\\
  \ind[B_f\notin E],&\alpha(f)=\mathsf{block}.
 \end{cases}\]
For $\tau=E_e$, let $m(\tau)=\mathsf{edge}$, and for
$\tau=B_e$, let $m(\tau)=\mathsf{block}$.  Let $\mathcal W_e^\tau$
contain the triples $(\pi,K,\alpha)$ such that $\pi$ is a simple
$s$-$t$ path, $K$ is an inclusion-minimal $s$-$t$ edge cut,
$\pi\cap K=\{e\}$, and $\alpha(e)=m(\tau)$.  The minimum over an empty family is $\Inf$.

\begin{theorem}\label{thm:path-responsibility}
Let $\tau\in\{E_e,B_e\}$.  Assume either that $\Ans(E)=1$ and
$\ell_E(\tau)$ is an enabling literal for $e$, or that $\Ans(E)=0$
and $\ell_E(\tau)$ is a disabling literal for $e$.  Then,
\[
 \Kappa_E(\tau)=
 \min_{(\pi,K,\alpha)\in\mathcal W_e^\tau}
 \bigl(a_E(\pi)+c_E(K,\alpha)-1\bigr).
 \tag{20}\label{eq:path-cut-responsibility}
\]
If the stated polarity condition fails, $\Kappa_E(\tau)=\Inf$.
\end{theorem}

The path and cut must intersect only at the candidate edge: any other common
edge contributes conflicting enabling and disabling literals.  At $e$, the
path and cut use complementary literals of the candidate; exactly one disagrees
with the observed state, which accounts for the subtraction of one.  All
remaining disagreements are contingency changes.  Theorem~\ref{thm:causal-characterization}
then gives equality with the minimum contingency size.

The cause criterion depends on potential paths, rather than only on the
observed derivation.  Consider potential paths
$\pi_1=(x_1,x_2)=s\to a\to t$ and
$\pi_2=(y_1,y_2)=s\to b\to t$.  Let $E$ contain the edge atoms for
$x_1,x_2,y_1$, omit the edge atom for $y_2$, and contain no blockers.
Only $\pi_1$ is active.  For $\tau=E_{y_1}$, choose
$\pi=\pi_2$, $K=\{x_1,y_1\}$, and let $\alpha$ select edge absence
on both cut edges.  Then,
\begin{enumerate}
    \item $a_E(\pi)=1,\, c_E(K,\alpha)=2\,\, \text{and}\,\,$
    \item $\Kappa_E(\tau)=1+2-1=2.$
\end{enumerate}

Thus, $E_{y_1}$ has responsibility $1/3$, although no active path uses
it.  The corresponding contingency deletes $E_{x_1}$ and inserts
$E_{y_2}$; the second path then sustains the answer, and toggling
$E_{y_1}$ destroys it.

Minimum cut \cite{ford1956maximal} computes true-outcome robustness;
Equation~\eqref{eq:path-radius} is shortest path with weights
$w_E(e)=\ind[E_e\notin E]+\ind[B_e\in E]\in\{0,1,2\}$.  For the directed-diamond family in
Corollary~\ref{cor:succinctness}, the potential graph is acyclic.  In a directed
acyclic graph, $e=(u,v)$ lies on a potential $s$-$t$ path exactly when
$u$ is reachable from $s$ and $t$ is reachable from $v$: a repeated
vertex in the concatenation would form a directed cycle.  Thus, two graph
searches decide actual-cause recognition for that family.

\begin{corollary}\label{cor:succinctness}
There is a fixed recursive stratified program and an explicitly
represented family $(I_n)_{n\geq1}$ of intervention instances of size
$O(n)$ such that, given $I_n$, a mutable state $F\subseteq\Univ_n$,
and a candidate atom $\tau\in\Univ_n$, the value $\Ans_{I_n}(F)$, the
cause status of $\tau$ at $E_n$, and $\Rob(E_n)$ are computable in time
polynomial in $|I_n|$, while $|\Prime_{I_n}^1|=\PowTwo{n}$.
\end{corollary}


\section{Complexity}\label{sec:complexity}
We construct a polynomial-size Boolean circuit for the response and
establish the complexity of the causal decision problems.

For a fixed safe stratified program and fixed goal predicate, with a
ground goal atom $A$ whose constants are part of the input, we construct circuit
$\Circ_{\PiP,\Dx,\Univ,A}$ in polynomial time by grounding $\PiP$ over
the active domain and unrolling each stratum.  The circuit has one input
$X_\tau$ for each
$\tau\in\Univ$. For $F\subseteq\Univ$, let $\ma X_F\in\{0,1\}^{\Univ}$ be its characteristic vector, defined by $\ma X_F(\tau)=1$ iff $\tau\in F$. Then,
\[
\forall F\subseteq\Univ,\quad\Circ_{\PiP,\Dx,\Univ,A}(\ma X_F)=\Ans(F),
 \tag{21}\label{eq:circuit}
\]
The circuit size is polynomial in $|\Dx|+|\Univ|+|A|$.  One gate represents each ground intensional atom at each immediate-consequence stage.  Lower strata supply completed gates to negated intensional body atoms.  Induction on stages and strata
proves \eqref{eq:circuit}.

For data complexity, the program $\PiP$ and the goal predicate are fixed, while
$\Dx,\Univ,E,A$ and, when applicable, $\tau\in\Univ$ and binary
$k\leq|\Univ|$ form the input.  The problems $\Cause$, $\Robust$, and
$\Responsibility$ ask whether $\tau$ is a cause, $\Rob(E)\leq k$, and
$\Kappa_E(\tau)\leq k$, respectively; equivalently,
$\Responsibility$ asks whether $\Resp_E(\tau)\geq 1/(k+1)$.

\begin{theorem}\label{thm:complexity}
For each problem below, there exists a fixed safe nonrecursive
stratified program for which the problem is NP-complete in data complexity.
\begin{enumerate}
\item $\Cause$, for an initially absent designated atom and a false
observed outcome.
\item $\Robust$, for a false observed outcome.
\item $\Responsibility$, for an initially absent designated atom.
\end{enumerate}
\end{theorem}

Membership guesses the contingency or intervention and evaluates the fixed
program on the required states.  The cause reduction uses exogenous facts
$\mathsf{Var}(x), \mathsf{Pos}(c,x)$, $\mathsf{Neg}(c,x)$, and
$\mathsf{Clause}(c)$, mutable assignment atoms $\mathsf{True}(x)$, and
an initially absent mutable atom $\mathsf{Switch}$.  One fixed program
contains
\[\begin{array}{rcl}
 \mathsf{Sat}(c)&\leftarrow&\mathsf{Pos}(c,x),\mathsf{True}(x),\\
 \mathsf{Sat}(c)&\leftarrow&\mathsf{Neg}(c,x),\mathsf{Var}(x), \neg\mathsf{True}(x),\\
 \mathsf{Bad}&\leftarrow&\mathsf{Clause}(c),\neg\mathsf{Sat}(c),\\
 \Goal&\leftarrow&\mathsf{Switch},\neg\mathsf{Bad}.
\end{array}
\tag{22}\label{eq:sat-program}\]
At the empty observed state, a contingency chooses precisely the variables
assigned true; negated body atoms interpret every other variable as false.
Before toggling $\mathsf{Switch}$, the goal remains false.  After the
toggle, the goal holds exactly when no clause derives $\mathsf{Bad}$, which
is equivalent to satisfaction of the encoded formula.  Thus, the designated
atom is a cause exactly for satisfiable instances.  For robustness, we reduce from $\VC$ restricted to nonempty graphs;
the restriction remains NP-complete by adding one disjoint edge and increasing
the threshold by one.  Exogenous edges and initially present mutable atoms
$\mathsf{Keep}(v)$ feed the fixed rules
\[\begingroup
\begin{array}{rcl}
 \mathsf{Bad}&\leftarrow&
 \mathsf{Edge}(u,v),\mathsf{Keep}(u),\mathsf{Keep}(v),\\[-0.5ex]
 \Goal&\leftarrow&\neg\mathsf{Bad}.
\end{array}
\tag{23}\label{eq:cover-program}
\endgroup\]
Initially, every vertex is kept, and $\mathsf{Bad}$ holds whenever the input
graph has an edge.  Deleting a family $S$ of keep-atoms removes
$\mathsf{Bad}$ exactly when every edge has an endpoint in $S$; hence,
$S$ is a vertex cover.  For responsibility, extend $\Univ$ by the
initially absent candidate $\tau=\mathsf{Switch}$; under
$\Goal\leftarrow\mathsf{Switch},\neg\mathsf{Bad}$, it becomes pivotal
precisely after deleting a vertex cover.  The minimum contingency size is
therefore the minimum cover size.  Each reduction uses a fixed nonrecursive stratified program and is
linear in the input size.  

The reductions also expose the quantitative meaning of the thresholds.
Consider the conjunctive normal form (CNF) formula $x\land y$, encoded by two unit clauses.  At the
empty mutable state, $\mathsf{Switch}$ is absent and the goal is false.
The contingency $\{\mathsf{True}(x),\mathsf{True}(y)\}$ satisfies both
clauses while preserving the false goal; inserting $\mathsf{Switch}$ then
derives the goal.  No smaller contingency satisfies the formula, so the
switch has minimum contingency size $2$ and responsibility $1/3$.
For robustness, let the exogenous graph be a triangle and let every
$\mathsf{Keep}$-atom be present.  Deleting one keep-atom leaves one edge
with both endpoints kept, whereas deleting any two removes every such edge.
Thus, the observed false goal has robustness radius $2$, equal to the
minimum vertex-cover size.  These instances separate polynomial evaluation of
a fixed rule program from the combinatorial search over admissible inputs.

For fixed programs
$\PiP_1, \PiP_2$ over a common extensional signature and designated goal
predicate, with possibly different auxiliary intensional predicates,
$\ResponseEquivalence$ takes common $(\Dx,\Univ,A)$.  Let $\Ans_j$ be the
response induced by $\PiP_j$; it asks whether,  $\forall F\subseteq\Univ,$
\[
 \quad\Ans_1(F)=\Ans_2(F).
\]

\begin{theorem}\label{thm:response-equivalence}
$\ResponseEquivalence$ is coNP-complete in data complexity for two fixed
nonrecursive stratified programs.
\end{theorem}

A nonequivalence certificate is one mutable state on which the two fixed
programs disagree.  For hardness, assume encoding a CNF formula by exogenous clause and positive- and
negative-occurrence relations and mutable truth-assignment atoms.  The first fixed program
derives the goal exactly on satisfying assignments; the second fixed
program is over the same declared signature and contains no rule with head
$\Goal$, thus it never derives the goal.  Their responses agree on all mutable states exactly when the formula
is unsatisfiable.  The construction is linear in the formula encoding.  Therefore,
even severe syntactic restrictions do not make universal preservation of
intervention behavior tractable.  Together with the separation at the observed state above, the theorem distinguishes
preservation at one observed state from preservation over the full intervention domain.
\section{Conclusion}
We have studied causal explanations for stratified Datalog under perfect-model
semantics and interventions over a finite set of mutable extensional facts.
We have shown that minimal supports and inclusion-minimal outcome-changing
interventions do not determine causality under negation. Using compatible prime
implicants for the observed and opposite outcomes, we have characterized minimum
contingencies, actual causes, responsibility, and robustness.
Our characterization recovers the positive-Datalog minimal-support theory, yields path–cut formulas for blocked recursive reachability, and characterizes
intervention-response preservation for replacements
over a fixed exogenous database, mutable domain, and goal.
Finally, we establish NP- and coNP-completeness results for fixed
nonrecursive stratified programs.

Several problems remain open. These include identifying tractable fragments, designing parameterized algorithms for bounded contingency size or treewidth, computing compatible prime pairs without explicit enumeration, and extending the semantics to undefined truth values and multiple models.

\paragraph*{Acknowledgments.}
This work is funded by the German Research Foundation (DFG) -- SFB 1574 Circular Factory-- 471687386.

\paragraph*{Declaration on the Use of Generative AI.}
GPT-5.6 Luna was used to improve language and readability. The authors reviewed and verified all AI-assisted edits and remain responsible for the content.
\clearpage
\bibliography{aaai2027}
\clearpage
\onecolumn
\appendix
\section*{Extended Proofs and Supplementary Results}\label{app:proofs}
\vspace{0.25cm}

\section{Supplementary Results}
We collect additional Boolean consequences, conservative specializations, and
solver reductions used by the main results.  

For a family $\Supp_I$ of minimal positive supports, consider the
standard monotone positive-support decomposition
\[
 \Ans(F)=1\quad\text{if and only if}\quad
 \exists S\in\Supp_I\;(S\subseteq F).
 \tag{25}\label{eq:support-decomposition}
\]
For transversals, we use $\Trans(\emptyset)=\{\emptyset\}$ and
$\Trans(\mathcal H)=\emptyset$ when $\emptyset\in\mathcal H$.

\begin{proposition}\label{prop:separation}
A fixed nonrecursive stratified program violates
\eqref{eq:support-decomposition}.  At the empty observed state, some atom is
an actual cause although it belongs to no inclusion-minimal intervention that
changes the observed outcome.
\end{proposition}

\begin{proposition}\label{prop:sign-necessity}
There is a fixed nonrecursive stratified program whose prime truth implicant
contains a negative literal and whose prime falsity implicant contains a
positive literal.
\end{proposition}

Let $\epsilon=(\emptyset,\emptyset)$,
$\mathbf 1=(\{\epsilon\},\emptyset)$, and
$\mathbf 0=(\emptyset,\{\epsilon\})$.  We use the identities
$\bigwedge\emptyset=1$ and $\bigvee\emptyset=0$; an empty rule body
therefore receives $\mathbf 1$, and an empty family of alternative rules
receives $\mathbf 0$.  For term families
$\mathcal C$ and $\mathcal D$, let $\Min_{\preceq}$ select their
$\preceq$-minimal members, and let
\[
 \mathcal C\otimes\mathcal D=
 \Min_{\preceq}\{C\merge D{\mid}C\in\mathcal C,\ D\in\mathcal D,
                    \ C,D\text{ compatible}\}.
\]
For pairs $X=(\mathcal T,\mathcal F)$ and
$Y=(\mathcal T',\mathcal F')$, let
\[
\begin{aligned}
 X\sqcap Y
  &=\bigl(\mathcal T\otimes\mathcal T',
          \Dual(\mathcal T\otimes\mathcal T')\bigr),\\
 X\primeor Y
  &=\bigl(\Dual(\mathcal F\otimes\mathcal F'),
          \mathcal F\otimes\mathcal F'\bigr),\\
 \neg X&=(\mathcal F,\mathcal T).
\end{aligned}
\tag{26}\label{eq:pair-operations}
\]
A mutable extensional atom $u$ receives
$(\{(\{u\},\emptyset)\},\{(\emptyset,\{u\})\})$; an exogenous true
atom receives $\mathbf1$; and every other extensional atom receives
$\mathbf0$.

For a ground intensional atom $H$, let $\Prime_I^b(H)$ denote the prime
implicants of
$F\mapsto\ind[H\in\PM(\PiP\cup\Dx\cup F)]$.  Ground $\PiP$ over the
active domain.  For a stratum $i$, let $m_i$ be its number of ground
intensional atoms and initialize $V_i^0(H)=\mathbf0$.  At stage $t+1$,
combine each ground rule body by $\sqcap$, combine alternative rules and
$V_i^t(H)$ by $\primeor$, use $V_i^t$ for positive atoms in stratum
$i$, use final pairs for lower-stratum atoms, and apply $\neg$ to negated
atoms.

For a Boolean response $f:2^{\Univ}\to\{0,1\}$, let
$\Prime^b(f)$ denote its prime $b$-implicants. The identities stated in Proposition~\ref{prop:calculus} are standard Boolean prime-implicant operations
\cite{crama2011boolean}.

\begin{proposition}\label{prop:calculus}
For Boolean responses $f$ and $g$ over
$\Univ$,
\begin{align}
 \Prime^1(f\land g)
 &=\Prime^1(f)\otimes\Prime^1(g),\notag\\
 \Prime^0(f\lor g)
 &=\Prime^0(f)\otimes\Prime^0(g),\notag\\
 \Prime^0(f\land g)
 &=\Dual(\Prime^1(f)\otimes\Prime^1(g)),\notag\\
 \Prime^1(f\lor g)
 &=\Dual(\Prime^0(f)\otimes\Prime^0(g)),\notag\\
 \Prime^b(\neg f)
 &=\Prime^{1-b}(f).
 \tag{27}\label{eq:prime-calculus}
\end{align}
\end{proposition}

\begin{proposition}\label{prop:relevance}
For $\tau\in\Univ$, the following conditions are equivalent:
\begin{enumerate}
\item some state $F$ satisfies
$\Ans(F)\neq\Ans(F\toggle\{\tau\})$;
\item {some member of $\Prime_I^1$ contains a literal on $\tau$};
\item {some member of $\Prime_I^0$ contains a literal on $\tau$}; and
\item $\tau$ is counterfactual at some state.
\end{enumerate}
\end{proposition}

Let $\Rel_I$ contain the atoms satisfying Proposition~\ref{prop:relevance}.
For every $F,J\subseteq\Univ$, 
\[\quad\Ans(F\toggle J)=\Ans(F\toggle(J\cap\Rel_I)).\]

\begin{corollary}\label{cor:invariance}
For intervention instances with common $\Dx$, $\Univ$, $E$, and $A$, equality of either
prime-implicant family is equivalent to response equality on every mutable
state.  Response equality implies equality of actual causes, responsibilities,
and robustness radii at $E$.
\end{corollary}

\begin{corollary}\label{cor:positive-polarity}
Assume that $\PiP$ is positive, and let $b=\Ans(E)$.
\begin{enumerate}
\item If $b=1$, every actual cause belongs to $E$, and
$\tau\in E$ is a cause exactly when it occurs in a minimal support; and
\item If $b=0$, every actual cause belongs to $\Univ\setminus E$, and
$\tau\notin E$ is a cause exactly when it occurs in a minimal transversal
of $\Supp_I$.
\end{enumerate}
\end{corollary}

For $\Pi_{\mathsf{br}}$ from Equation~\eqref{eq:blocked-path}, {an atom
whose literal satisfied by a true state enables its edge is counterfactual
exactly when every active $s$-$t$ path uses that edge. An atom whose
literal satisfied by a false state disables its edge is counterfactual exactly
when, on some potential path, the complementary activation literal of the
candidate is the only activation literal not satisfied by the state.}

For a fixed program and fixed goal predicate, $\Implicant$ takes
$(\Dx,\Univ,A,C,b)$, where $C=(P,N)$ satisfies
$P,N\subseteq\Univ$ and $P\cap N=\emptyset$, and where
$b\in\{0,1\}$; it asks whether $C$ is a $b$-implicant.  For two fixed
programs over a common extensional signature and common designated goal
predicate, $\ResponseEquivalence$ asks whether their responses agree on every
state over a common mutable domain.

\begin{proposition}\label{prop:circuit}
For every fixed safe stratified program and fixed goal predicate, the
ground goal atom $A$, including its constants, is input.  A polynomial-time
construction maps $(\Dx,\Univ,A)$ to a Boolean circuit
$\Circ_{\PiP,\Dx,\Univ,A}$, with one input $X_\tau$ for each
$\tau\in\Univ$, such that $\forall F\subseteq\Univ,$
\[
 \quad \Circ_{\PiP,\Dx,\Univ,A}(\chi_F)=\Ans(F),
\]
where $\chi_F$ is the characteristic function of $F$.  The circuit size is
polynomial in $|\Dx|+|\Univ|+|A|$.
\end{proposition}

Let $e_u=\chi_E(u)$.  For Boolean values $x,y \in \{0,1\}$, 
$x \oplus y = 1$ exactly when $x \neq y$. Let $X^{\oplus\tau}$ be obtained from
$X$ by complementing coordinate $\tau$.

\begin{proposition}\label{prop:solver-encoding}
Let $\Circ=\Circ_{\PiP,\Dx,\Univ,A}$ and $b=\Ans(E)$.
\begin{enumerate}
\item $\Rob(E)\leq k$ exactly when
\[
 \Circ(X)=1-b\,\, \,\text{and}\,\,\,
 \sum_{u\in\Univ}(X_u\oplus e_u)\leq k
 \tag{28}\label{eq:robust-encoding}
\]
is satisfiable; and
\item $\Kappa_E(\tau)\leq k$ exactly when
\[
\begin{gathered}
 X_\tau=e_\tau,
 \quad \Circ(X)=b,
 \quad \Circ(X^{\oplus\tau})=1-b\quad\text{and}\quad
 \sum_{u\in\Univ\setminus\{\tau\}}(X_u\oplus e_u)\leq k
\end{gathered}
\tag{29}\label{eq:responsibility-encoding}
\]
is satisfiable.  Removing the cardinality constraint decides $\Cause$.
\end{enumerate}
\end{proposition}

\begin{proposition}\label{prop:profile-separation}
There are two fixed nonrecursive stratified Datalog programs over a common
$(\Dx,\Univ,E,A)$ with the same actual causes, responsibilities, and
robustness radius at $E$, but different responses on $2^{\Univ}$.
\end{proposition}

\begin{theorem}\label{thm:implicant-complexity}
$\Implicant$ is coNP-complete in data complexity for a fixed nonrecursive
stratified program.  Hardness holds for the empty term and outcome $0$.
\end{theorem}

\vspace{0.25cm}

\section{Extended Proofs of Main Results}
\vspace{0.25cm}

\subsubsection*{Proof of Theorem~\ref{thm:abstraction-separation}}
We compute both responses on all mutable states, derive their support and
baseline-change families, and then determine  minimum
contingency of every candidate at the observed state.

{Let $\PiP_1$ be the program in Equation~\eqref{eq:separation-program},
let $\PiP_2=\{\Goal\leftarrow P(c)\}$, and let
\[
 I_1=(\PiP_1,\emptyset,\{p,q,r\},\emptyset,\Goal)
  \quad\text{and}\quad
 I_2=(\PiP_2,\emptyset,\{p,q,r\},\emptyset,\Goal).
\]}
For every $F\subseteq\{p,q,r\}$,
\[
\begin{aligned}
 \Ans_{I_1}(F)=1
 &\Longleftrightarrow
 (p\in F\land q\notin F\land r\notin F)
 \lor(p\in F\land q\in F\land r\in F),\\
 \Ans_{I_2}(F)=1
 &\Longleftrightarrow p\in F.
\end{aligned}
\tag{31}\label{eq:two-responses}
\]

\emph{Minimal supports.}
For each instance, $\Ans(\{p\})=1$ and $\Ans(\emptyset)=0$, thus
$\{p\}$ is a minimal support.  Every true state of either instance contains
$p$.  Therefore, no support omitting $p$ exists, and every support
properly containing $\{p\}$ is nonminimal.  Thus,
\[
 \Supp_{I_1}=\Supp_{I_2}=\{\{p\}\}.
\tag{32}\label{eq:same-support}
\]

\emph{Minimal outcome changes.}
The observed state is $E=\emptyset$, and both observed outcomes are $0$.
For $I_1$, the true intervention states are $\{p\}$ and
$\{p,q,r\}$; the latter properly contains the former.  For $I_2$, every
true intervention state contains $p$, and $\{p\}$ is true.  Thus,
\[
 \Flip_{I_1}(E)=\Flip_{I_2}(E)=\{\{p\}\}.
\tag{33}\label{eq:same-flips}
\]

\emph{Candidate $p$.}
For both instances, $\Ans(E)=0$ and $\Ans(E\toggle\{p\})=1$.
Thus, $p$ is counterfactual,
$\Kappa_E^{I_1}(p)=\Kappa_E^{I_2}(p)=0$, and both responsibility values are
$1$.

\emph{Candidate $q$ in $I_1$.}
Let $\Gamma_q=\{p,r\}$.  Equation~\eqref{eq:two-responses} gives
\[
 \Ans_{I_1}(E\toggle\Gamma_q)=\Ans_{I_1}(\{p,r\})=0,\quad\text{and}\quad
 \Ans_{I_1}(E\toggle(\Gamma_q\cup\{q\}))
 =\Ans_{I_1}(\{p,q,r\})=1.
\]
Thus, $\Kappa_E^{I_1}(q)\leq2$.  The contingencies excluding $q$ of
size at most one are $\emptyset$, $\{p\}$, and $\{r\}$.  Their
adjacent outcome pairs before and after toggling $q$ are, respectively,
$(0,0), (1,0)$ and $(0,0)$.
None satisfies~\eqref{eq:cause}.  Therefore,
$\Kappa_E^{I_1}(q)=2$ and $\Resp_E^{I_1}(q)=1/3$.

\emph{Candidate $r$ in $I_1$.}
Let $\Gamma_r=\{p,q\}$.  Then,
\[
 \Ans_{I_1}(E\toggle\Gamma_r)=\Ans_{I_1}(\{p,q\})=0, \quad\text{and}\quad
 \Ans_{I_1}(E\toggle(\Gamma_r\cup\{r\}))
 =\Ans_{I_1}(\{p,q,r\})=1.
\]
The contingencies excluding $r$ of size at most one are
$\emptyset$, $\{p\}$, and $\{q\}$, with adjacent outcome pairs
$(0,0), (1,0) $ and $(0,0)$.
Thus, $\Kappa_E^{I_1}(r)=2$ and $\Resp_E^{I_1}(r)=1/3$.

\emph{Candidates $q$ and $r$ in $I_2$.}
The second response depends only on $p$.  For every contingency excluding
$q$, toggling $q$ preserves membership of $p$, and therefore preserves
the outcome.  For every contingency excluding $r$, toggling $r$ also preserves membership of $p$, and therefore preserves the outcome.
\[
 \Kappa_E^{I_2}(q)=\Kappa_E^{I_2}(r)=\Inf,
 \,\,\text{and}\,\,
 \Resp_E^{I_2}(q)=\Resp_E^{I_2}(r)=0.
\]
Equations~\eqref{eq:same-support} and~\eqref{eq:same-flips}, together with the
computed responsibility values, prove the theorem.

\subsubsection*{Proof of Theorem~\ref{thm:remote-contingency}}
For each n, we derive the response of the fixed program, identify all inclusion-minimal outcome-changing interventions, and prove
matching upper and lower bounds on each designated contingency.

For $n\geq2$, let
\[
 \Dx=\{\mathsf{Item}(a_i){\mid}1\leq i\leq n\} \quad \text{and}\quad
 \Univ=\{g\}\cup\{s_i{\mid}1\leq i\leq n\},
\]
where $s_i=\mathsf{Sel}(a_i)$, and let $E=\emptyset$.  Use the fixed
program~\eqref{eq:remote-program}.  It is safe and nonrecursive.  A
stratification places extensional predicates in stratum $0$,
$\mathsf{Some}$ and $\mathsf{Missing}$ in stratum $1$, and
$\mathsf{None}$, $\mathsf{All}$, and $\Goal$ in stratum $2$.

For a state $F\subseteq\Univ$, let
$S_F=F\cap\{s_1,\ldots,s_n\}$.  Perfect-model evaluation gives
\[
\begin{aligned}
 \mathsf{Some}\in\PM(\PiP\cup\Dx\cup F)
 &\Longleftrightarrow S_F\neq\emptyset,\\
 \mathsf{Missing}\in\PM(\PiP\cup\Dx\cup F)
 &\Longleftrightarrow S_F\neq\{s_1,\ldots,s_n\},\\
 \Ans(F)=1
 &\Longleftrightarrow
 g\in F\ \land\
 \bigl(S_F=\emptyset\ \lor\ S_F=\{s_1,\ldots,s_n\}\bigr).
\end{aligned}
\tag{36}\label{eq:remote-response}
\]
The observed outcome is $0$.  Equation~\eqref{eq:remote-response} gives
$\Ans(\{g\})=1$, thus $\Rob(E)\leq1$.  No intervention of size $0$
changes the outcome, hence $\Rob(E)=1$.

Let $J$ be an inclusion-minimal outcome-changing intervention.  Since
$E=\emptyset$, its reached state is $J$, and
Equation~\eqref{eq:remote-response} gives $g\in J$.  If
$S_J=\emptyset$, then $J=\{g\}$.  If
$S_J=\{s_1,\ldots,s_n\}$, then $\{g\}\subsetneq J$ already has
outcome $1$, contrary to inclusion-minimality.  Thus,
\[
 \Flip_I(E)=\{\{g\}\}.
\tag{37}\label{eq:remote-flip}
\]

For $i\in\{1,\ldots,n\}$, let
$\Gamma_i=\{g\}\cup\{s_j{\mid}j\neq i\}$.
The reached state has a nonempty proper selection family, thus
$\Ans(E\toggle\Gamma_i)=0$.  After toggling $s_i$, every selection is
present and Equation~\eqref{eq:remote-response} gives outcome $1$.  Hence,
$s_i$ is a cause and $\Kappa_E(s_i)\leq|\Gamma_i|=n$.

For the reverse inequality, let $\Gamma$ be any contingency witnessing that
$s_i$ is a cause.  Since $s_i\notin E$ and
$s_i\notin\Gamma$, the state $E\toggle\Gamma$ omits $s_i$, while
$E\toggle(\Gamma\cup\{s_i\})$ contains it.  The latter state has outcome
$1$.  By Equation~\eqref{eq:remote-response}, it must contain $g$, and
its selection family cannot be empty because it contains $s_i$.  Therefore,
it contains every $s_j$.  Thus,
\[
 \{g\}\cup\{s_j{\mid}j\neq i\}\subseteq\Gamma,
 \qquad
 |\Gamma|\geq n.
\]
We obtain $\Kappa_E(s_i)=n$ and
$\Resp_E(s_i)=1/(n+1)$ for every $i$.  The database and mutable domain
have size $O(n)$, while the program is fixed.

\subsubsection*{Proof of Proposition~\ref{prop:representation}}
{
Let $b\in\{0,1\}$.  We prove the prime cover, the duality identity, and
response determination.

\emph{Prime cover and forward direction.}
Assume $\Ans(F)=b$.  The complete term
$C_F=(F,\Univ\setminus F)$
has exactly one satisfying state.  Indeed, $G\models C_F$ implies
$F\subseteq G$ and $G\cap(\Univ\setminus F)=\emptyset$; hence,
$G\subseteq F$ and $G=F$.  Thus, $C_F$ is a $b$-implicant.
The family of subterms of $C_F$ has cardinality at most
$2^{|F|}2^{|\Univ\setminus F|}$ and is finite.  The nonempty family of
$b$-implicant subterms therefore has a $\preceq$-minimal member $C$.
By Definition~\ref{def:prime}, $C\in\Prime_I^b$.  Since
$C\preceq C_F$ and $F\models C_F$, we have $F\models C$.

\emph{Prime cover and reverse direction.}
Assume $F\models C$ for some $C\in\Prime_I^b$.  The implicant condition
in Definition~\ref{def:prime} applies to every satisfying state, in particular
to $F$.  Hence, $\Ans(F)=b$.  The two directions establish
\eqref{eq:prime-cover}.

\emph{Duality, inclusion $\Prime_I^{1-b}\subseteq\Dual(\Prime_I^b)$.}
Let $D\in\Prime_I^{1-b}$.  For every $C\in\Prime_I^b$, the terms $C$
and $D$ conflict.  If they were compatible, their merged term would admit a
state $F$ satisfying both, and the two implicant conditions would give
$\Ans(F)=b$ and $\Ans(F)=1-b$, a contradiction.

Let
\[
 \mathcal S_D=\{D'\preceq D\mid
   D'\text{ conflicts with every }C\in\Prime_I^b\}.
\]
The family is nonempty because $D\in\mathcal S_D$, and it is finite because
$D$ has finitely many subterms.  Choose a $\preceq$-minimal
$D_0\in\mathcal S_D$.  We show that $D_0$ is a $(1-b)$-implicant.  Let
$F\models D_0$.  If $\Ans(F)=b$, the prime cover yields
$C\in\Prime_I^b$ with $F\models C$.  The common satisfying state $F$
would make $C$ and $D_0$ compatible, contradicting
$D_0\in\mathcal S_D$.  Thus, $\Ans(F)=1-b$ for every
$F\models D_0$.

We now have a $(1-b)$-implicant $D_0\preceq D$.  Since $D$ is prime,
$D_0=D$.  The minimality of $D_0$ in $\mathcal S_D$ is exactly the
minimality required by $\Dual(\Prime_I^b)$, thus
$D\in\Dual(\Prime_I^b)$.

\emph{Duality, inclusion $\Dual(\Prime_I^b)\subseteq\Prime_I^{1-b}$.}
Let $D\in\Dual(\Prime_I^b)$, and let $F\models D$.  If
$\Ans(F)=b$, the prime cover yields $C\in\Prime_I^b$ with $F\models C$,
contradicting the defining conflict condition for $D$.  Therefore,
$\Ans(F)=1-b$, and $D$ is a $(1-b)$-implicant.

Among the finitely many $(1-b)$-implicant subterms of $D$, choose a prime
one $D_0\preceq D$.  The inclusion proved above gives
$D_0\in\Dual(\Prime_I^b)$.  Since $D$ is minimal among terms conflicting
with every member of $\Prime_I^b$, we have $D\preceq D_0$.  Together with
$D_0\preceq D$, antisymmetry gives $D=D_0$, and hence
$D\in\Prime_I^{1-b}$.  Equation~\eqref{eq:prime-duality} follows.

\emph{Response determination.}
Let instances $I_1$ and $I_2$ share $(\Dx,\Univ,A)$.  If
$\Prime_{I_1}^1=\Prime_{I_2}^1$, then for every $F\subseteq\Univ$,
\[
\begin{aligned}
 \Ans_{I_1}(F)=1
 &\Longleftrightarrow
 \exists C\in\Prime_{I_1}^1\;(F\models C)\\
 &\Longleftrightarrow
 \exists C\in\Prime_{I_2}^1\;(F\models C)\\
 &\Longleftrightarrow \Ans_{I_2}(F)=1.
\end{aligned}
\]
Since both responses are Boolean, equality of their truth values implies full
response equality.  If $\Prime_{I_1}^0=\Prime_{I_2}^0$, then for every
$F\subseteq\Univ$,
\[
\begin{aligned}
 \Ans_{I_1}(F)=0
 &\Longleftrightarrow
 \exists C\in\Prime_{I_1}^0\;(F\models C)\\
 &\Longleftrightarrow
 \exists C\in\Prime_{I_2}^0\;(F\models C)\\
 &\Longleftrightarrow \Ans_{I_2}(F)=0.
\end{aligned}
\]
Booleanity again gives full response equality.  Conversely, response equality
makes the implicant predicate identical for every term and every polarity; the
common subterm order then makes primeness identical.  Thus, both prime families
are equal.  Duality also shows that equality of either polarity entails
equality of the other.
}

\subsubsection*{Proof of Proposition~\ref{prop:compositional}}
Recall the stratification thus that all intensional predicates have positive
strata and the used intensional strata are $1,\ldots,s$.  For a mutable state
$F\subseteq\Univ$, let $M_{i,F}^t$ be the family of ground intensional
atoms in stratum $i$ obtained after $t$ applications of the inflationary
immediate-consequence operator for that stratum.  Extensional facts and the
completed relations of lower strata remain constant during that iteration.
For a ground stratum-$i$ atom $H$, let
\[
 h_{H,t}(F)=1
 \Longleftrightarrow
 H\in M_{i,F}^t.
\]
We prove, simultaneously for every stratum-$i$ atom $H$,
\[
 V_i^t(H)=\bigl(\Prime^1(h_{H,t}),\Prime^0(h_{H,t})\bigr)
 \tag{38}\label{eq:stage-invariant}
\]
by outer induction on $i$ and inner induction on $t$.

\emph{Outer induction .}
For every $j<i$ and every ground atom $L$ in stratum $j$, the final pair
computed for $L$ equals the prime pair of its completed perfect-model
response.  For $i=1$, no lower intensional stratum exists, thus the hypothesis
is vacuous.

\emph{Inner base case.}
For every $F$, $M_{i,F}^0=\emptyset$.  Thus, $h_{H,0}$ is the
constant-zero response.  Its prime $1$-family is empty, and its unique prime
$0$-implicant is the empty term $\epsilon$.  Therefore,
\[
 V_i^0(H)=\mathbf0
 =\bigl(\emptyset,\{\epsilon\}\bigr)
 =\bigl(\Prime^1(h_{H,0}),\Prime^0(h_{H,0})\bigr).
\]

\emph{Inner induction .}
Assume~\eqref{eq:stage-invariant} for every ground atom in stratum $i$ at
stage $t$.  Let $\rho$ be a ground stratum-$i$ rule with head $H$, and
let $b_{\rho,t}$ be its Boolean body response when positive stratum-$i$
atoms are interpreted by $M_{i,F}^t$.

We first prove the exact pair for the body by induction on its length.  For an
empty body, the response is the constant-one function.  Its unique prime
$1$-implicant is $\epsilon$, and its prime $0$-family is empty, thus the
assigned pair is $\mathbf1$.

Assume that the first $k$ body atoms have response $g_k$ and exact pair
$(\Prime^1(g_k),\Prime^0(g_k))$.  Let $L$ be the $(k+1)$-st body atom,
and let $\lambda_L$ be its Boolean response.  We verify the exact pair for
$L$ in every syntactically possible case.

\begin{enumerate}
\item If $L=u$ is a mutable extensional atom, then
$\lambda_L(F)=1$ exactly when $u\in F$.  Its prime families are
$\{(\{u\},\emptyset)\}$ and
$\{(\emptyset,\{u\})\}$, which are the initialized pair.

\item If $L$ is an exogenous extensional atom, its value is independent of
$F$.  When the atom belongs to $\Dx$, its response is constant one and its
pair is $\mathbf1$.  Otherwise, its response is constant zero and its pair is
$\mathbf0$.

\item If $L$ is a positive atom in a lower intensional stratum, the outer
induction hypothesis supplies its exact completed pair.

\item If $L$ is a positive atom in stratum $i$, the inner induction
hypothesis supplies the exact pair of its stage-$t$ response.

\item If $L=\neg u$ for an extensional atom $u$, Cases 1 and 2 give the
exact pair for $u$.  Proposition~\ref{prop:calculus} exchanges its two prime
families under negation, exactly as the operation $\neg X$ in
\eqref{eq:pair-operations}.

\item If $L=\neg K$ for an intensional atom $K$, stratification places
$K$ in a stratum below $i$.  The outer induction hypothesis gives the
completed exact pair for $K$, and Proposition~\ref{prop:calculus} gives the
exact swapped pair for $\neg K$.
\end{enumerate}

The six cases are exhaustive because a safe stratified rule body contains
positive or negated extensional atoms, positive current- or lower-stratum
intensional atoms, and negated lower-stratum intensional atoms; stratification
excludes negated current- and higher-stratum intensional atoms.  The response
of the first $k+1$ body atoms is $g_k\land\lambda_L$.  By
Proposition~\ref{prop:calculus}, applying $\sqcap$ to their exact pairs yields
\[
 \bigl(\Prime^1(g_k\land\lambda_L),
       \Prime^0(g_k\land\lambda_L)\bigr).
\]
Induction on the body length gives
\[
 \mathsf{Pair}_{\rho,t}
 =\bigl(\Prime^1(b_{\rho,t}),\Prime^0(b_{\rho,t})\bigr).
 \tag{39}\label{eq:body-pair}
\]

Let $R_i(H)=\{\rho_1,\ldots,\rho_m\}$ be the ground stratum-$i$ rules with
head $H$.  Let
\[
 q_0=h_{H,t}\quad\text{and}\quad
 q_j=q_{j-1}\lor b_{\rho_j,t}
 \quad(1\leq j\leq m).
\]
The exact pair for $q_0$ follows from the inner induction hypothesis.  If the
iteration has the exact pair for $q_{j-1}$, then
Equation~\eqref{eq:body-pair} and the disjunction identity in
Proposition~\ref{prop:calculus} show that $\primeor$ produces the exact pair
for $q_j$.  Induction on $j$ yields the exact pair for
\[
 q_m=h_{H,t}\lor\bigvee_{\rho\in R_i(H)}b_{\rho,t}
     =h_{H,t+1}.
 \tag{40}\label{eq:stage-response}
\]
The case $R_i(H)=\emptyset$ has $m=0$ and gives
$h_{H,t+1}=h_{H,t}$, so it is included.  Consequently,
\[
 V_i^{t+1}(H)
 =\bigl(\Prime^1(h_{H,t+1}),\Prime^0(h_{H,t+1})\bigr),
\]
which proves the inner induction step.

Let $m_i$ be the number of ground intensional atoms in stratum $i$.  For
each state $F$, the sequence
$M_{i,F}^0\subseteq M_{i,F}^1\subseteq\cdots$
is inflationary.  Every strict inclusion adds at least one of the $m_i$
available atoms.  Hence, no sequence has more than $m_i$ strict increases,
and
\[
 M_{i,F}^{m_i}=M_{i,F}^{m_i+1}
\]
for every $F$.  The stabilized interpretation is the least fixpoint of the
positive immediate-consequence operator for stratum $i$.  Applying
\eqref{eq:stage-invariant} at stage $m_i$ gives
\[
 V_i^{m_i}(H)=\bigl(\Prime_I^1(H),\Prime_I^0(H)\bigr).
\]
Thus, the final values in stratum $i$ satisfy the outer induction claim.
Induction over $i=1,\ldots,s$ proves the proposition.

\subsubsection*{Proof of Theorem~\ref{thm:causal-characterization}}
Let $b=\Ans(E)$ and $\tau\in\Univ$.  We prove each item in both
directions.  The responsibility proof constructs a contingency from a
compatible prime pair and constructs a compatible prime pair from a minimum
contingency.

The candidate polarity cases are exhaustive.  If $\tau\in E$, then
$\ell_E(\tau)=\tau$ and $\ell_E^{\mathsf c}(\tau)=\neg\tau$.  If
$\tau\notin E$, then $\ell_E(\tau)=\neg\tau$ and
$\ell_E^{\mathsf c}(\tau)=\tau$.  Each alternative occurs with $b=1$ and
with $b=0$.  The derivations below use only these complementary literals and
therefore discharge all four cases.

\emph{Item 1, forward direction.}
Assume that $\tau$ is an actual cause, and let $\Gamma$ satisfy
\eqref{eq:cause}.  Let
\[
 F=E\toggle\Gamma\,\,\,\,\text{and}\,\,\,
 G=F\toggle\{\tau\}.
\]
Since $\tau\notin\Gamma$, the states $E$ and $F$ assign the same
value to $\tau$.  Equation~\eqref{eq:cause} gives
$\Ans(F)=b$ and $\Ans(G)=1-b$.  By~\eqref{eq:prime-cover}, choose
$C\in\Prime_I^b$ with $F\models C$.  {If $C$ contained no literal on $\tau$, then}
$G\models C$, and the implicant property would give $\Ans(G)=b$.  Thus,
{$C$ contains a literal on $\tau$.  Since $F$ agrees with $E$ on $\tau$, that literal is
$\ell_E(\tau)$.}

\emph{Item 1, reverse direction.}
Assume that $C\in\Prime_I^b$ contains {$\ell_E(\tau)$}.  Let
$C_{-\tau}$ remove {that literal}.  Primeness implies that
$C_{-\tau}$ is not a $b$-implicant.  Thus, a state $G$ satisfies
\[
 G\models C_{-\tau} \,\,\,\,\text{and}\,\,\,
 \Ans(G)=1-b.
\tag{41}\label{eq:prime-witness}
\]
The state $G$ assigns $\tau$ the value opposite to its current value.  If
it assigned the current value, then $G\models C$, contradicting
\eqref{eq:prime-witness}.  Let $F=G\toggle\{\tau\}$.  Then,
$F\models C$, $\Ans(F)=b$, and $F$ agrees with $E$ on $\tau$.
Let $\Gamma=E\toggle F$.  We have
\[
 \tau\notin\Gamma,
 \quad
 E\toggle\Gamma=F \,\,\,\,\text{and}\,\,\,
 E\toggle(\Gamma\cup\{\tau\})=G.
\]
These equalities and~\eqref{eq:prime-witness} establish~\eqref{eq:cause}.

\emph{Item 2, upper bound.}
Let $(C,D)\in\mathcal Q_{E,\tau}^b$, and let
$M=C\merge D^\tau=(P_M,N_M)$.
Compatibility makes $M$ a term.  {Both components contain
$\ell_E(\tau)$.}  Let
\[
 F=(E\cup P_M)\setminus N_M \,\,\,\,\text{and}\,\,\,
 \Gamma=E\toggle F.
\]
{For every positive literal of $M$, $F$ contains its atom; for every
negative literal, $F$ omits its atom.  Thus, $F\models M$.  Since
$\ell_E(\tau)$ is satisfied by $E$, $\tau\notin\Gamma$.  The remaining
literals of $M$ not satisfied by $E$ correspond exactly to the
noncandidate members of $\Gamma$, thus}
\[
 |\Gamma|=\Dist_{E,-\tau}(M).
\tag{42}\label{eq:constructed-distance}
\]
Since $F\models C$, $\Ans(F)=b$.  Since $F\models D^\tau$, {toggling}
$\tau$ yields $F\toggle\{\tau\}\models D$, and therefore
$\Ans(F\toggle\{\tau\})=1-b$.  Therefore, $\Gamma$ is a contingency, and
\[
 \Kappa_E(\tau)
 \leq\Dist_{E,-\tau}(C\merge D^\tau).
\]
Minimization over $\mathcal Q_{E,\tau}^b$ gives
$\Kappa_E(\tau)\leq K_E(\tau)$.

\emph{Item 2, lower bound.}
Assume $\Kappa_E(\tau)<\Inf$, and let $\Gamma$ be a contingency of
minimum size.  Let
\[
 F=E\toggle\Gamma \,\,\,\,\text{and}\,\,\,
 G=F\toggle\{\tau\}.
\]
Choose $C\in\Prime_I^b$ with $F\models C$, and choose
$D\in\Prime_I^{1-b}$ with $G\models D$.  The forward direction of
Item~1 shows that $C$ contains {$\ell_E(\tau)$}.  {If $D$ contained no literal on $\tau$, then} $F\models D$, contradicting $\Ans(F)=b$.  {The
literal on $\tau$ in $D$ agrees with $G$, thus it is
$\ell_E^{\mathsf c}(\tau)$.}
Hence, $F\models D^\tau$.  The common satisfying state $F$ proves that
$C$ and $D^\tau$ are compatible, and therefore
$(C,D)\in\mathcal Q_{E,\tau}^b$.

Let $M=C\merge D^\tau=(P_M,N_M)$.  Since $F=E\toggle\Gamma$ satisfies
$M$, every atom in $(P_M\setminus E)\setminus\{\tau\}$ belongs to
$F\setminus E\subseteq\Gamma$, and every atom in
$(N_M\cap E)\setminus\{\tau\}$ belongs to
$E\setminus F\subseteq\Gamma$.  The two atom families are disjoint.  Thus,
\[
 \Dist_{E,-\tau}(M)\leq|\Gamma|=\Kappa_E(\tau).
\]
Taking the minimum over compatible pairs yields
$K_E(\tau)\leq\Kappa_E(\tau)$.  The two bounds give
$K_E(\tau)=\Kappa_E(\tau)$.  Equation~\eqref{eq:responsibility} gives the
responsibility formula.  If $\mathcal Q_{E,\tau}^b=\emptyset$, the
lower-bound construction proves that no contingency exists; both minima are
$\Inf$.

\emph{Item 3, upper bound.}
Let $D=(P,N)\in\Prime_I^{1-b}$, and let
\[
 F=(E\cup P)\setminus N.
\]
Then, $F\models D$ and $\Ans(F)=1-b$.  The toggled atoms are exactly
$P\setminus E$ and $N\cap E$, thus
$|E\toggle F|=\Dist_E(D)$.  Therefore,
\[
 \Rob(E)\leq\Dist_E(D).
\]
Minimization over $D$ gives the upper bound in~\eqref{eq:robustness-prime}.

\emph{Item 3, lower bound.}
Assume $\Rob(E)<\Inf$, and choose $J\subseteq\Univ$ with
$|J|=\Rob(E)$ and $\Ans(E\toggle J)=1-b$.  By~\eqref{eq:prime-cover},
choose $D=(P,N)\in\Prime_I^{1-b}$ satisfied by $E\toggle J$.  Every atom
in $P\setminus E$ must be inserted by $J$, and every atom in
$N\cap E$ must be deleted by $J$.  These families are disjoint.  Thus,
\[
 \Dist_E(D)=|P\setminus E|+|N\cap E|
 \leq|J|=\Rob(E).
\]
The upper and lower bounds prove~\eqref{eq:robustness-prime}.  If
$\Prime_I^{1-b}=\emptyset$, Equation~\eqref{eq:prime-cover} states that no
opposite-outcome state exists, so both sides are $\Inf$.

\subsubsection*{Proof of Proposition~\ref{prop:compatibility-necessity}}
{
Let $p=P(c)$, $q=Q(c)$, $r=R(c)$,
$\Univ=\{p,q,r\}$, and $E=\{p,r\}$.  {Let $\PiP_3$ be the program
in Equation~\eqref{eq:compatibility-example} and let
$I_3=(\PiP_3,\emptyset,\Univ,E,\Goal)$.}  The program is ground, safe,
nonrecursive, and stratified.
For every $F\subseteq\Univ$,
\[
 \Ans(F)=1
 \Longleftrightarrow
 r\notin F\ \land\ (p\notin F\ \lor\ q\notin F).
 \tag{43}\label{eq:compatibility-response}
\]
Since $r\in E$, the observed outcome is $0$.

We compute the prime $0$-family.  The term
$C_r=(\{r\},\emptyset)$ forces outcome $0$.  Its only proper subterm is
$\epsilon$, which is satisfied by $\emptyset$ of outcome $1$; hence,
$C_r$ is prime.  The term $C_{pq}=(\{p,q\},\emptyset)$ also forces outcome
$0$: when $r\in F$, the first conjunct in
\eqref{eq:compatibility-response} fails, and when $r\notin F$, the two
disjuncts fail because $p,q\in F$.  Removing $p$ yields
$(\{q\},\emptyset)$, which is satisfied by $\{q\}$ of outcome $1$.
Removing $q$ yields $(\{p\},\emptyset)$, which is satisfied by
$\{p\}$ of outcome $1$.  Thus, $C_{pq}$ is prime.

Every false state either contains $r$ or contains both $p$ and $q$.  If
a $0$-implicant contained neither $r$ nor both $p,q$ as positive
requirements, it would admit a true state obtained by omitting $r$ and one
of $p,q$.  Therefore, every prime $0$-implicant equals $C_r$ or
$C_{pq}$, and
\[
 \Prime_I^0=\{(\{r\},\emptyset),(\{p,q\},\emptyset)\}.
 \tag{44}\label{eq:compatibility-prime-zero}
\]

For outcome $1$, let
$D_p=(\emptyset,\{p,r\})$ and $ D_q=(\emptyset,\{q,r\})$.
Every state satisfying either term omits $r$ and omits one of $p,q$, thus it
has outcome $1$.  In $D_p$, removing $\neg r$ admits $\{r\}$ of
outcome $0$, and removing $\neg p$ admits $\{p,q\}$ of outcome $0$.
Thus, $D_p$ is prime.  The symmetric witnesses $\{r\}$ and $\{p,q\}$
show that both literals of $D_q$ are necessary.  Every true state omits
$r$ and at least one of $p,q$, thus every prime $1$-implicant equals
$D_p$ or $D_q$.  Therefore,
\[
 \Prime_I^1=\{(\emptyset,\{p,r\}),(\emptyset,\{q,r\})\}.
 \tag{45}\label{eq:compatibility-prime-one}
\]

For candidate $p$, $\ell_E(p)=p$.  The only prime opposite-outcome term
containing $\neg p$ is $D_p$.  Outside the candidate coordinate, its sole
literal is $\neg r$, which is not satisfied by $E$.  Hence,
\[
 \OppDist_E(p)=\Dist_{E,-p}(D_p)=1.
 \tag{46}\label{eq:compatibility-naive-distance}
\]

Let $\Gamma=\{q,r\}$.  Then,
\[
 E\toggle\Gamma=\{p,q\}
  \quad\text{and}\quad
 E\toggle(\Gamma\cup\{p\})=\{q\},
\]
and~\eqref{eq:compatibility-response} gives outcomes $0$ and $1$,
respectively.  Therefore, $p$ is a cause and $\Kappa_E(p)\leq2$.

The contingencies excluding $p$ of size at most one are exactly
$\emptyset$, $\{q\}$, and $\{r\}$.  Their adjacent outcome pairs are

\[
\begin{array}{|c|c|c|}
\hline
 \Gamma & \Ans(E\toggle\Gamma)
        & \Ans(E\toggle(\Gamma\cup\{p\}))\\
\hline
 \emptyset & \Ans(\{p,r\})=0 & \Ans(\{r\})=0\\
\hline
 \{q\} & \Ans(\{p,q,r\})=0 & \Ans(\{q,r\})=0\\
\hline
 \{r\} & \Ans(\{p\})=1 & \Ans(\emptyset)=1\\
\hline
\end{array}
\]
No row satisfies Definition~\ref{def:cause}.  Thus,
$\Kappa_E(p)\geq2$, and $\Kappa_E(p)=2$.

Finally, the current prime term $C_{pq}$ contains $p$.  Replacing
$\neg p$ in $D_p$ by $p$ gives
$D_p^p=(\{p\},\{r\})$, which is compatible with $C_{pq}$.  Their merge is
$(\{p,q\},\{r\})$.  Outside $p$, the literals not satisfied by $E$ are
$q$ and $\neg r$, thus Equation~\eqref{eq:pair-distance} returns $2$.
The compatible current-outcome term accounts for the contingency change that
the opposite-only distance omits.
}

\subsubsection*{Proof of Corollary~\ref{cor:rewrite}}
{
Let $I$ and $I'$ share $\Dx$, $\Univ$, and $A$.

Assume that the replacement is intervention-preserving.  Then,
$\Ans_I(F)=\Ans_{I'}(F)$ for every $F\subseteq\Univ$.  For every term
$C$ and polarity $b$,
\[
\begin{aligned}
 C\text{ is a }b\text{-implicant for }I
 &\Longleftrightarrow
 \forall F\models C,\ \Ans_I(F)=b\\
 &\Longleftrightarrow
 \forall F\models C,\ \Ans_{I'}(F)=b\\
 &\Longleftrightarrow
 C\text{ is a }b\text{-implicant for }I'.
\end{aligned}
\]
The subterm order is common to the two instances.  Hence, $C$ has a proper
$b$-implicant subterm for $I$ exactly when it has one for $I'$.  Thus,
\[
 \Prime_I^b=\Prime_{I'}^b
 \qquad(b=0,1).
\]

Conversely, assume $\Prime_I^1=\Prime_{I'}^1$.  For every mutable state
$F$, the prime cover gives
\[
\begin{aligned}
 \Ans_I(F)=1
 &\Longleftrightarrow
 \exists C\in\Prime_I^1\;(F\models C)\\
 &\Longleftrightarrow
 \exists C\in\Prime_{I'}^1\;(F\models C)\\
 &\Longleftrightarrow
 \Ans_{I'}(F)=1.
\end{aligned}
\]
The Boolean responses therefore agree on every state, so the replacement is
intervention-preserving.  If $\Prime_I^0=\Prime_{I'}^0$, then for every
$F\subseteq\Univ$,
\[
\begin{aligned}
 \Ans_I(F)=0
 &\Longleftrightarrow \exists C\in\Prime_I^0\;(F\models C)\\
 &\Longleftrightarrow \exists C\in\Prime_{I'}^0\;(F\models C)\\
 &\Longleftrightarrow \Ans_{I'}(F)=0.
\end{aligned}
\]
The Boolean responses agree on every mutable state.  Equation~\eqref{eq:prime-duality}
also shows that equality of the prime $1$-families is equivalent to equality
of the prime $0$-families.  Both directions are proved.
}

\subsubsection*{Proof of Corollary~\ref{cor:positive}}
{
Assume that $\PiP$ is positive.  Perfect-model entailment is monotone in the
mutable state:
\[
 F\subseteq F'
 \Longrightarrow
 \Ans(F)\leq\Ans(F').
 \tag{50}\label{eq:monotonicity}
\]

\emph{Prime truth implicants are minimal supports.}
Let $S\in\Supp_I$.  Every state $F$ satisfying
$(S,\emptyset)$ contains $S$.  Since $\Ans(S)=1$, monotonicity gives
$\Ans(F)=1$.  Thus, $(S,\emptyset)$ is a $1$-implicant.  For every
$s\in S$, minimality of $S$ gives $\Ans(S\setminus\{s\})=0$, and the
state $S\setminus\{s\}$ satisfies the term obtained by deleting the literal
$s$.  Hence, every literal is necessary and $(S,\emptyset)$ is prime.

Conversely, let $C=(P,N)\in\Prime_I^1$.  The state $P$ satisfies $C$
because $P\cap N=\emptyset$, thus $\Ans(P)=1$.  Every state containing $P$
has outcome $1$ by monotonicity; therefore, $(P,\emptyset)$ is a
$1$-implicant and a subterm of $C$.  Primeness gives $N=\emptyset$.  If
a proper subset $P'\subsetneq P$ had outcome $1$, monotonicity would make
$(P',\emptyset)$ a proper $1$-implicant subterm of $C$, a
contradiction.  Thus, $P$ is a minimal support.  We obtain
\[
 \Prime_I^1=\{(S,\emptyset)\mid S\in\Supp_I\}.
\]

\emph{Prime falsity implicants are minimal transversals.}
Let $T\in\Trans(\Supp_I)$, and let $F\models(\emptyset,T)$.  Then,
$F\cap T=\emptyset$.  Suppose $\Ans(F)=1$.  Among the finitely many
substates of $F$ with outcome $1$, choose an inclusion-minimal one $S$.
Then, $S\in\Supp_I$ and $S\subseteq F$, thus
$S\cap T=\emptyset$, contradicting that $T$ intersects every support.
Therefore, $(\emptyset,T)$ is a $0$-implicant.

Let $t\in T$.  Since $T$ is an inclusion-minimal transversal,
$T\setminus\{t\}$ is not a transversal.  Hence, some
$S_t\in\Supp_I$ satisfies
\[
 S_t\cap(T\setminus\{t\})=\emptyset.
\]
Since $T$ is a transversal, $S_t\cap T\neq\emptyset$; the preceding
equality forces $S_t\cap T=\{t\}$.  Thus, $S_t$ satisfies the term
obtained from $(\emptyset,T)$ by deleting $\neg t$, and
$\Ans(S_t)=1$.  Every negative literal is necessary, thus
$(\emptyset,T)$ is prime.

Conversely, let $C=(P,N)\in\Prime_I^0$.  We first show that
$(\emptyset,N)$ is a $0$-implicant.  Suppose
$F\cap N=\emptyset$ and $\Ans(F)=1$.  Since $P\cap N=\emptyset$, the
state $F\cup P$ also avoids $N$.  Monotonicity gives
$\Ans(F\cup P)=1$, while $F\cup P\models(P,N)=C$, contradicting the
$0$-implicant property.  Thus, $(\emptyset,N)$ is a $0$-implicant.
Because it is a subterm of the prime term $C$, we obtain $P=\emptyset$.

The family $N$ intersects every support.  Otherwise, a support
$S\in\Supp_I$ with $S\cap N=\emptyset$ would satisfy
$(\emptyset,N)$ and have outcome $1$.  If a proper subset
$N'\subsetneq N$ intersected every support, the argument in the preceding
paragraph would show that $(\emptyset,N')$ is a proper $0$-implicant
subterm, contradicting primeness.  Hence, $N$ is an inclusion-minimal
transversal, and
\[
 \Prime_I^0=\{(\emptyset,T)\mid T\in\Trans(\Supp_I)\}.
\]

The boundary conventions also agree.  If $\Supp_I=\emptyset$, no state has
outcome $1$, thus $\Prime_I^1=\emptyset$, the response is constant zero,
and $\Prime_I^0=\{\epsilon\}$; our convention gives
$\Trans(\emptyset)=\{\emptyset\}$.  If $\emptyset\in\Supp_I$,
monotonicity makes the response constant one, thus
$\Prime_I^1=\{\epsilon\}$ and $\Prime_I^0=\emptyset$; our convention
gives $\Trans(\Supp_I)=\emptyset$.  Thus, both equalities in
\eqref{eq:positive-specialization} hold in all cases.
}

\subsubsection*{Proof of Theorem~\ref{thm:path-cut}}
We prove the path and cut representations in both directions, derive the two robustness formulas by matching upper and lower bounds, and then prove the cause and succinctness clauses.

Let $H=(V,L)$, and use the notation from Section~\ref{sec:reachability}.

\emph{Item 1, forward direction.}
Let $\pi$ be a simple $s$-$t$ path.  Every state satisfying $C_\pi$
contains $E_e$ and omits $B_e$ for every $e\in\pi$.  Thus, every edge of
$\pi$ is active, and $\Pi_{\mathsf{br}}$ derives
$\Path(s,t)$.

To prove primeness, remove one {literal} from $C_\pi$.  If the removed
literal is the {positive literal} $E_e$, choose a state containing exactly the edge
facts for $\pi\setminus\{e\}$, omitting all blockers on $\pi$, and omitting
all other edge facts.  If the removed literal is the {negative literal}
$\neg B_e$, choose a state containing all edge facts on $\pi$, containing
$B_e$, omitting the other blockers on $\pi$, and omitting all other edge
facts.  In either case, the reduced term is satisfied and the active graph has
no $s$-$t$ path.  Thus, {every literal is necessary}, and $C_\pi$ is prime.

\emph{Item 1, reverse direction.}
Let $C=(P,N)\in\Prime_I^1$.
{
Suppose $B_e\in P$.  Remove the positive blocker literal $B_e$, and let
$F$ satisfy the reduced term.  If $B_e\in F$, then $F\models C$ and has
an active $s$-$t$ path.  If $B_e\notin F$, let
$F'=F\cup\{B_e\}$.  The state $F'$ satisfies $C$, so it has an active
$s$-$t$ path.  Passing from $F'$ to $F$ removes a blocker and can only activate
edge $e$; hence, every path active in $F'$ remains active in $F$. Thus, the reduced term is a $1$-implicant, contradicting primeness.  Thus,
$P$ contains no blocker atom.

Suppose $E_e\in N$, thus $C$ contains the negative edge literal
$\neg E_e$.  Remove that literal, and let $F$ satisfy the reduced term.  If
$E_e\notin F$, then $F\models C$ and has an active path.  If
$E_e\in F$, let $F'=F\setminus\{E_e\}$.  The state $F'$ satisfies
$C$, so it has an active path.  Passing from $F'$ to $F$ inserts an edge
fact and can only activate edge $e$; the path remains active.  Again, the
reduced term is a $1$-implicant, contradicting primeness.  Hence, $N$
contains no edge atom.
}
Primeness therefore gives
\[
 P\subseteq\{E_e{\mid}e\in L\} \,\,\,\,\text{and}\,\,\,
 N\subseteq\{B_e{\mid}e\in L\}.
 \tag{51}\label{eq:truth-types}
\]

Let $F_C$ contain exactly the edge facts in $P$, omit exactly the blockers
in $N$, omit every other edge fact, and contain every other blocker.  The
state $F_C$ satisfies $C$.  Since $C$ is a $1$-implicant, the active
graph $H_{F_C}$ contains a simple $s$-$t$ path $\pi$.  Every edge
$e\in\pi$ satisfies $E_e\in P$ and $B_e\in N$.  Hence,
$C_\pi\preceq C$.  The forward direction proved that $C_\pi$ is a
$1$-implicant.  Primeness of $C$ gives $C=C_\pi$.

\emph{Item 2, forward direction.}
Let $K$ be an inclusion-minimal directed $s$-$t$ edge cut, and let
$\alpha:K\to\{\mathsf{edge},\mathsf{block}\}$.  Every state satisfying
$D_{(K,\alpha)}$ disables each edge of $K$: an edge fact is absent when
$\alpha(e)=\mathsf{edge}$, and a blocker is present when
$\alpha(e)=\mathsf{block}$.  Its active graph is a subgraph of $H-K$, thus
it has no $s$-$t$ path.  Thus, $D_{(K,\alpha)}$ is a $0$-implicant.

Let $e\in K$.  Inclusion-minimality of $K$ gives a directed
$s$-$t$ path $\pi_e$ in $H-(K\setminus\{e\})$.  Since $K$ is a
cut, $\pi_e$ uses $e$; thus, $\pi_e\cap K=\{e\}$.  Remove the {literal}
selected for $e$ from $D_{(K,\alpha)}$, and construct state $F_e$ as
follows.  For every $f\in\pi_e$, let $E_f\in F_e$ and
$B_f\notin F_e$.  For every $f\in K\setminus\{e\}$, let
$E_f\notin F_e$ when $\alpha(f)=\mathsf{edge}$, and let
$B_f\in F_e$ when $\alpha(f)=\mathsf{block}$.  These assignments do not
conflict because $\pi_e\cap(K\setminus\{e\})=\emptyset$.  For every edge
not yet assigned, let its edge atom be absent and its blocker atom be present.
The resulting state satisfies every retained literal of the reduced term.  It
also contains every edge atom and omits every blocker on $\pi_e$, thus it
activates $\pi_e$.  Thus, removing
any selected {literal} admits a true state.  {Every selected literal is necessary},
and $D_{(K,\alpha)}$ is prime.

\emph{Item 2, reverse direction.}
Let $D=(P,N)\in\Prime_I^0$.
{
Suppose $E_e\in P$.  Remove the positive edge literal $E_e$, and let
$F$ satisfy the reduced term.  If $E_e\in F$, then $F\models D$ and has
no active $s$-$t$ path.  If $E_e\notin F$, let
$F'=F\cup\{E_e\}$.  The state $F'$ satisfies $D$ and has no active path.
Passing from $F'$ to $F$ deletes an edge fact and can only remove active
edges, thus $F$ also has no active path.  The reduced term is a $0$-implicant,
contradicting primeness.  Hence, $P$ contains no edge atom.

Suppose $B_e\in N$, so $D$ contains the negative blocker literal
$\neg B_e$.  Remove that literal, and let $F$ satisfy the reduced term.  If
$B_e\notin F$, then $F\models D$ and has no active path.  If
$B_e\in F$, let $F'=F\setminus\{B_e\}$.  The state $F'$ satisfies
$D$ and has no active path.  Passing from $F'$ to $F$ inserts a blocker
and can only remove active edge $e$, so $F$ also has no active path.  The
reduced term is a $0$-implicant, contradicting primeness.  Hence, $N$
contains no blocker atom.
}
Thus,
\[
 P\subseteq\{B_e{\mid}e\in L\} \,\,\,\,\text{and}\,\,\,\,
 N\subseteq\{E_e{\mid}e\in L\}.
 \tag{52}\label{eq:false-types}
\]
{
No edge contributes both $B_e\in P$ and $E_e\in N$.  If both occurred,
remove either one.  The remaining literal still disables $e$, and all other
edge constraints are unchanged.  Every state satisfying the reduced term still
disables every edge disabled by the original term, thus it remains unreachable.
The reduced term would be a proper $0$-implicant, contradicting primeness.
}

Let
\[
 K=\{e{\mid}B_e\in P\text{ or }E_e\in N\}.
\]
If $K$ were not a directed $s$-$t$ cut, choose a directed path
$\pi$ with $\pi\cap K=\emptyset$.  Construct a state $F$ by putting
$E_f\in F$ and $B_f\notin F$ for every $f\in\pi$; putting
$B_f\in F$ when $B_f\in P$; putting $E_f\notin F$ when
$E_f\in N$; and assigning all remaining edge facts absent and all remaining
blockers present.  The path avoids $K$, so these assignments are consistent.
The state $F$ satisfies $D$ and activates $\pi$, contradicting the
$0$-implicant property.  Thus, $K$ is a cut.  If a proper subset
$K'\subsetneq K$ were a cut, retain in $D$ only the selected {literal} for
each edge of $K'$.  Every state satisfying that proper subterm disables all
edges in $K'$, and therefore has no directed $s$-$t$ path.  The subterm
would be a proper $0$-implicant, contradicting primeness.  Thus, $K$ is
inclusion-minimal, and $D=D_{(K,\alpha)}$ for the unique mode map determined
by its {literals}.

\emph{Item 3.}
Assume $H_E$ contains an $s$-$t$ path.  Let $J$ change the outcome to
false.  Let $X_J$ contain the edges active in $H_E$ that are inactive in
$H_{E\toggle J}$.  Every $s$-$t$ path in $H_E$ intersects $X_J$;
otherwise, that path would remain active after {applying $J$}.  Thus, $X_J$
is an $s$-$t$ edge cut of $H_E$.  Making an active edge inactive requires
toggling its present edge fact or its absent blocker, and distinct edges require
distinct facts.  Therefore,
\[
 |J|\geq|X_J|\geq\lambda_{H_E}(s,t).
 \tag{53}\label{eq:cut-lower}
\]

Conversely, let $K$ be a minimum edge cut of $H_E$.  For every $e\in K$,
toggle either the present edge fact $E_e$ or the absent blocker $B_e$.  The
resulting intervention has size $|K|$, only disables active edges, and removes
every active $s$-$t$ path.  Thus,
$\Rob(E)\leq|K|=\lambda_{H_E}(s,t)$.  Together with
\eqref{eq:cut-lower}, the two bounds prove \eqref{eq:cut-radius}.

\emph{Item 4.}
Assume $H_E$ has no $s$-$t$ path.  If $H$ has no potential
$s$-$t$ path, no intervention can create the goal, and both sides of
\eqref{eq:path-radius} equal $\Inf$.  Assume that $H$ has a potential path.
For a potential path $\pi$, let
\[
 J_\pi=\{E_e{\mid}e\in\pi,E_e\notin E\}
 \cup
 \{B_e{\mid}e\in\pi,B_e\in E\}.
\]
{Applying the intervention $J_\pi$} makes every edge of $\pi$ active.  Its cardinality is the
sum in \eqref{eq:path-radius}, thus the minimum path cost is an upper bound on
$\Rob(E)$.

Conversely, let $J$ create the goal, and choose an active $s$-$t$ path
$\pi$ in $H_{E\toggle J}$.  For every $e\in\pi$, if $E_e\notin E$,
then $E_e\in J$; if $B_e\in E$, then $B_e\in J$.  {These literals concern distinct mutable atoms.}  Therefore,
\[
 |J|\geq
 \sum_{e\in\pi}
 \bigl(\ind[E_e\notin E]+\ind[B_e\in E]\bigr).
\]
Minimization over $J$ and $\pi$ gives the lower bound and establishes
\eqref{eq:path-radius}.

\emph{Item 5.}
{
Assume first that $\Ans(E)=1$.  A prime $1$-implicant is a path term
$C_\pi$ by Item 1.  Such a term contains only the enabling literals
$E_f$ and $\neg B_f$ for edges $f\in\pi$.

Let $\tau=E_e$.  If $E_e\in E$, then
$\ell_E(\tau)=E_e$, and $C_\pi$ contains that literal exactly when
$e\in\pi$.  If $E_e\notin E$, then
$\ell_E(\tau)=\neg E_e$, and no path term contains that disabling literal.
Let $\tau=B_e$.  If $B_e\notin E$, then
$\ell_E(\tau)=\neg B_e$, and $C_\pi$ contains it exactly when
$e\in\pi$.  If $B_e\in E$, then
$\ell_E(\tau)=B_e$, and no path term contains it.  By Item 1 of
Theorem~\ref{thm:causal-characterization}, a candidate is therefore a cause in
a true state exactly when its current literal enables $e$ and $e$ lies on
a simple potential $s$-$t$ path.

Assume next that $\Ans(E)=0$.  A prime $0$-implicant is a cut term
$D_{(K,\alpha)}$ by Item 2.  Such a term contains only disabling literals:
$\neg E_f$ in edge mode and $B_f$ in blocker mode.

For $\tau=E_e$, the current literal occurs in a cut term exactly when
$E_e\notin E$, $e\in K$, and $\alpha(e)=\mathsf{edge}$.  For
$\tau=B_e$, it occurs exactly when $B_e\in E$, $e\in K$, and
$\alpha(e)=\mathsf{block}$.  Hence, the causal characterization reduces the
false-outcome cause condition to membership of $e$ in an inclusion-minimal
directed $s$-$t$ edge cut.

We prove that an edge belongs to an inclusion-minimal directed
$s$-$t$ cut exactly when it belongs to a directed $s$-$t$ path.  Let
$e\in K$ for an inclusion-minimal cut $K$.  Since
$K\setminus\{e\}$ is not a cut, the graph
$H-(K\setminus\{e\})$ contains an $s$-$t$ path $\pi$.  The path must
use $e$, because $K$ is a cut.  Thus, $e\in\pi$.

Conversely, let $\pi$ be an $s$-$t$ path containing $e$.  The family
$K_0=\{e\}\cup(L\setminus\pi)$ is a cut: every $s$-$t$ path either is
$\pi$, in which case it uses $e$, or contains an edge outside $\pi$.
Among the cuts $K'\subseteq K_0$ that contain $e$, choose an
inclusion-minimal one $K$.  For every $f\in K\setminus\{e\}$, the family
$K\setminus\{f\}$ still contains $e$; its being a cut would contradict the
choice of $K$.  The family $K\setminus\{e\}$ is not a cut because the path
$\pi$ avoids $K_0\setminus\{e\}$, and hence avoids
$K\setminus\{e\}$.  Thus, deleting any member of $K$ destroys the cut
property, thus $K$ is inclusion-minimal and contains $e$.  The false-outcome
cause characterization follows.

\emph{Counterfactual cases.}
{Suppose $\Ans(E)=1$ and $\ell_E(\tau)$ is an enabling literal for $e$.
We distinguish the two possible edge states.  If $e$ is inactive in $H_E$,
then toggling $\tau$ cannot remove any active edge and therefore cannot destroy
an active $s$-$t$ path.  Thus, $\tau$ is not counterfactual.  Since
$\Ans(E)=1$, an active $s$-$t$ path exists, and no active path uses the inactive
edge $e$; therefore, the condition that every active $s$-$t$ path uses $e$ is
also false.  If $e$ is active in $H_E$, toggling $\tau$ disables exactly $e$
and changes no other edge status.  The resulting state is false exactly when
every active $s$-$t$ path in $H_E$ uses $e$.  Hence, in both cases, $\tau$ is
counterfactual exactly when every active $s$-$t$ path uses $e$.}

Suppose $\Ans(E)=0$ and the current literal of $\tau$ disables $e$.
Toggling $\tau$ can activate only $e$.  If the toggle creates an active
$s$-$t$ path $\pi$, then $e\in\pi$, the candidate activation literal
is the only activation literal of $\pi$ not satisfied by $E$, and every
other edge of $\pi$ is already active.  Conversely, if a potential path
$\pi$ has the complementary candidate activation literal as its only
unsatisfied activation literal, toggling $\tau$ satisfies that literal and
activates every edge of $\pi$.  The outcome changes from $0$ to $1$.
These two implications prove the auxiliary criterion.
}

\emph{Item 6.}
{
Let $H_n=(V_n,L_n)$, where
\[
 V_n=\{a_0,\ldots,a_n,u_1,\ldots,u_n,v_1,\ldots,v_n\}
\]
and, for every $1\leq i\leq n$, $L_n$ contains exactly
$(a_{i-1},u_i),(u_i,a_i),(a_{i-1},v_i),(v_i,a_i)$.  Let
$s=a_0$ and $t=a_n$.  In layer $i$, every
$a_{i-1}$-$a_i$ path uses exactly one of
\[
 a_{i-1}\to u_i\to a_i
 \qquad\text{and}\qquad
 a_{i-1}\to v_i\to a_i.
\]
Since consecutive layers meet only at $a_i$, every simple
$a_0$-$a_n$ path chooses one branch in each layer, and every vector in
$\{0,1\}^n$ determines one such path.  This is a bijection, thus $H_n$ has
exactly {$\PowTwo{n}$} simple $s$-$t$ paths.  Distinct paths have distinct edge
families and therefore distinct terms $C_\pi$.  Item 1 gives
{$|\Prime_{I_{H_n,E_n}}^1|=\PowTwo{n}$} for
$E_n=\emptyset$.
}

\subsubsection*{Proof of Theorem~\ref{thm:path-responsibility}}
Let $\tau\in\{E_e,B_e\}$, and let $m(\tau)$ be its cut mode.

\emph{True observed outcome.}
Assume $\Ans(E)=1$ and {$\ell_E(\tau)$ enables $e$}.  If
$\tau=E_e$, then $E_e\in E$, the path term contains $E_e$, and a cut
term with $\alpha(e)=\mathsf{edge}$ contains $\neg E_e$.  If
$\tau=B_e$, then $B_e\notin E$, the path term contains $\neg B_e$, and
a cut term with $\alpha(e)=\mathsf{block}$ contains $B_e$.  Therefore,
every pair in $\mathcal Q_{E,\tau}^1$ has the form
\[
 (C_\pi,D_{(K,\alpha)}),
\]
where $e\in\pi\cap K$ and $\alpha(e)=m(\tau)$.

Let $f\in(\pi\cap K)\setminus\{e\}$.  If
$\alpha(f)=\mathsf{edge}$, then $C_\pi$ contains $E_f$ and
$D_{(K,\alpha)}^\tau$ contains $\neg E_f$.  If
$\alpha(f)=\mathsf{block}$, then $C_\pi$ contains $\neg B_f$ and
$D_{(K,\alpha)}^\tau$ contains $B_f$.  In both cases, the terms conflict.
At $e$, {replacing the candidate literal in the cut term by its complement
makes it equal to the path literal}.  Edges outside $\pi\cap K$ occur in at most one term.
Therefore,
\[
 C_\pi\text{ is compatible with }D_{(K,\alpha)}^\tau
 \Longleftrightarrow
 \pi\cap K=\{e\}.
\tag{54}\label{eq:path-cut-compatibility}
\]
Thus, $\mathcal Q_{E,\tau}^1$ is in bijection with
$\mathcal W_e^\tau$.

Let $(\pi,K,\alpha)\in\mathcal W_e^\tau$. Each
$f\in\pi\setminus\{e\}$ contributes
$\ind[E_f\notin E]+\ind[B_f\in E]$ to the distance. Each
$f\in K\setminus\{e\}$ contributes $\ind[E_f\in E]$ in edge mode and
$\ind[B_f\notin E]$ in blocker mode. These contributions concern distinct
atoms because $\pi\cap K=\{e\}$. {At $e$, we distinguish the two
candidate atoms. If $\tau=E_e$, then $E_e\in E$; the candidate cut literal
$\neg E_e$ contributes $1$ to $c_E(K,\alpha)$ and is removed by switching and
by $\Dist_{E,-\tau}$, while the noncandidate path literal $\neg B_e$
contributes $\ind[B_e\in E]$ to both sides. If $\tau=B_e$, then $B_e\notin E$;
the candidate cut literal $B_e$ contributes $1$ to $c_E(K,\alpha)$; switching
it changes the literal to $\neg B_e$, and $\Dist_{E,-\tau}$ excludes the
candidate coordinate. The noncandidate path literal $E_e$ contributes
$\ind[E_e\notin E]$ to both sides. Thus, the subtraction of one cancels
exactly the candidate cut cost and retains the other activation cost at $e$.}
Hence,
\[
 \Dist_{E,-\tau}(C_\pi\merge D_{(K,\alpha)}^\tau)
 =a_E(\pi)+c_E(K,\alpha)-1.
\tag{55}\label{eq:path-cut-distance}
\]
Item~2 of Theorem~\ref{thm:causal-characterization}, together with
\eqref{eq:path-cut-compatibility} and~\eqref{eq:path-cut-distance}, proves the
formula for a true observed outcome.

\emph{False observed outcome.}
Assume $\Ans(E)=0$ and {$\ell_E(\tau)$ disables $e$}.  If
$\tau=E_e$, then $E_e\notin E$, the cut term in edge mode contains
$\neg E_e$, and the path term contains $E_e$.  If $\tau=B_e$, then
$B_e\in E$, the cut term in blocker mode contains $B_e$, and the path
term contains $\neg B_e$.  Every pair in $\mathcal Q_{E,\tau}^0$ has the
form
\[
 (D_{(K,\alpha)},C_\pi),
\]
where $e\in\pi\cap K$ and $\alpha(e)=m(\tau)$.

For every $f\in(\pi\cap K)\setminus\{e\}$, the cut and path {literals are complementary} under both cut modes, thus they
conflict.  At $e$, {replacing the candidate literal in $C_\pi$ by its
complement makes it equal to the cut literal $\ell_E(\tau)$.}
Therefore,
\[
 D_{(K,\alpha)}\text{ is compatible with }C_\pi^\tau
 \Longleftrightarrow
 \pi\cap K=\{e\}.
\tag{55a}\label{eq:path-cut-compatibility-false}
\]
{For a compatible triple, the path and cut literals outside $e$ contribute
the activation and disabling costs stated above. At $e$, we again distinguish
the two candidates. If $\tau=E_e$, then $E_e\notin E$; the candidate path
literal $E_e$ contributes $1$ to $a_E(\pi)$ and is removed by switching and by
$\Dist_{E,-\tau}$, while the noncandidate path literal $\neg B_e$ contributes
$\ind[B_e\in E]$ to both sides. If $\tau=B_e$, then $B_e\in E$; the
candidate path literal $\neg B_e$ contributes $1$ to $a_E(\pi)$; switching
it changes the literal to $B_e$, and $\Dist_{E,-\tau}$ excludes the candidate
coordinate. The noncandidate path literal $E_e$ contributes
$\ind[E_e\notin E]$ to both sides. Thus, the subtraction of one cancels
exactly the candidate path cost and retains the other activation cost at $e$.}
Hence,
\[
 \Dist_{E,-\tau}(D_{(K,\alpha)}\merge C_\pi^\tau)
 =a_E(\pi)+c_E(K,\alpha)-1.
\tag{55b}\label{eq:path-cut-distance-false}
\]
Item~2 of Theorem~\ref{thm:causal-characterization} proves the formula for a
false observed outcome.

\emph{Failed polarity.}
{Assume $\Ans(E)=1$ and $\ell_E(\tau)$ disables $e$.  Every prime
$1$-implicant is a path term, and path terms contain only enabling literals.
Thus, no prime implicant of the observed outcome contains $\ell_E(\tau)$.
Assume $\Ans(E)=0$ and $\ell_E(\tau)$ enables $e$.  Every prime
$0$-implicant is a cut term, and cut terms contain only disabling literals.
Again, no prime implicant of the observed outcome contains $\ell_E(\tau)$.}  Item~1 of Theorem~\ref{thm:causal-characterization} gives
that $\tau$ is not a cause in either case, and
$\Kappa_E(\tau)=\Inf$.

\subsubsection*{Proof of Corollary~\ref{cor:succinctness}}
{
Let $H_n$ be the graph from Item 6 and let
$I_n=I_{H_n,E_n}$, where $E_n=\emptyset$.  Every potential edge atom
and blocker atom is mutable.  The graph has $3n+1$ vertices, $4n$
potential edges, and $8n$ mutable atoms, thus $|I_n|=O(n)$.  The program
$\Pi_{\mathsf{br}}$ is independent of $n$.

Given an explicitly represented $I_n$ and a mutable state
$F\subseteq\Univ_n$, perfect-model evaluation is polynomial in
$|I_n|$ because $\Pi_{\mathsf{br}}$ is fixed; more concretely, it
computes directed reachability in the active subgraph induced by $F$.

We next give a polynomial cause-recognition procedure.  At $E_n=\emptyset$,
the outcome is false.  For an edge atom $E_e$, the current literal is
$\neg E_e$, which disables $e$.  By Item 5 of
Theorem~\ref{thm:path-cut}, $E_e$ is a cause exactly when $e$ belongs to a
potential $s$-$t$ path.  The potential graph is acyclic.  For
$e=(u,v)$, such a path exists exactly when $u$ is reachable from $s$ and
$t$ is reachable from $v$.  Two graph searches compute these conditions
for every edge in polynomial time.  For a blocker atom $B_e$, the current
literal is $\neg B_e$, which enables rather than disables $e$.  The failed
polarity clause of Theorem~\ref{thm:path-responsibility} gives that no blocker
atom is a cause at the false state $E_n$.  Thus, actual-cause recognition is
polynomial for all candidate atoms.

The robustness radius is polynomially computable.  At $E_n$, every edge fact
is absent and every blocker is absent, so the false-outcome formula
\eqref{eq:path-radius} assigns cost $1$ to each edge of a potential path.
Each potential edge has weight
$w_{E_n}(e)=\ind[E_e\notin E_n]+\ind[B_e\in E_n]=1$, thus a shortest-path
computation returns the value in Equation~\eqref{eq:path-radius}.  More
generally, the false-outcome weight is
$w_E(e)=\ind[E_e\notin E]+\ind[B_e\in E]\in\{0,1,2\}$, while the
true-outcome case uses Equation~\eqref{eq:cut-radius}; both are computable in
polynomial time.

Finally, each diamond contributes one independent binary branch choice.
Mapping a binary vector in $\{0,1\}^n$ to its selected branches is a bijection
with the simple $s$-$t$ paths.  Hence, there are exactly
{$\PowTwo{n}$} paths.
Item 1 of Theorem~\ref{thm:path-cut} maps each path to a distinct prime truth
term and maps every prime truth term back to a path.  Therefore,
\[
 {|\Prime_{I_n}^1|=\PowTwo{n}.}
\]
Thus, given $I_n$, $F\subseteq\Univ_n$, and
$\tau\in\Univ_n$, the value $\Ans_{I_n}(F)$, the cause status of
$\tau$ at $E_n$, and $\Rob(E_n)$ are computable in time polynomial
in $|I_n|$, while the prime truth family has $\PowTwo{n}$ members.
}

\subsubsection*{Proof of Theorem~\ref{thm:complexity}}
{
We use exactly the programs and predicate arities stated in
Equations~\eqref{eq:sat-program} and~\eqref{eq:cover-program}.  Each item has a
fixed safe nonrecursive stratified program and a fixed goal predicate.

\emph{Membership in NP.}
For $\Cause$, guess
$\Gamma\subseteq\Univ\setminus\{\tau\}$, evaluate the fixed program on
$E\toggle\Gamma$ and
$E\toggle(\Gamma\cup\{\tau\})$, and verify~\eqref{eq:cause}.  For
$\Robust$, guess $J\subseteq\Univ$ with $|J|\leq k$, evaluate the
program on $E$ and $E\toggle J$, and verify opposite outcomes.  For
$\Responsibility$, guess
$\Gamma\subseteq\Univ\setminus\{\tau\}$ with $|\Gamma|\leq k$ and
verify~\eqref{eq:cause}.  Each witness has polynomial size, and evaluation of a
fixed stratified program is polynomial in the database-level input.  Thus, all
three problems belong to NP.

\emph{Item 1: hardness of $\Cause$.}
Let $\varphi$ be a CNF formula with variable family $V$ and clause family
$K$.  Let $\Dx$ contain
\[
 \mathsf{Var}(v),\quad \mathsf{Clause}(c),\quad
 \mathsf{Pos}(c,v)\quad\text{and}\quad \mathsf{Neg}(c,v)
\]
for the variables, clauses, positive occurrences, and negative occurrences of
$\varphi$.  Let
\[
 \Univ=\{\mathsf{True}(v)\mid v\in V\}\cup\{\mathsf{Switch}\},
 \quad E=\emptyset,
 \quad A=\Goal \quad\text{and}\quad \tau=\mathsf{Switch}.
\]
Use the fixed program in Equation~\eqref{eq:sat-program}.  It is safe and
nonrecursive.  A stratification places the extensional predicates in stratum
$0$, $\mathsf{Sat}$ in stratum $1$, $\mathsf{Bad}$ in stratum $2$,
and $\Goal$ in stratum $3$.

For a contingency
$\Gamma\subseteq\{\mathsf{True}(v)\mid v\in V\}$, let
$\nu_\Gamma(v)=1$ exactly when $\mathsf{True}(v)\in\Gamma$.  For every
clause $c$, the first two rules of~\eqref{eq:sat-program} give
\[
 \mathsf{Sat}(c)\in\PM(\PiP\cup\Dx\cup\Gamma)
 \Longleftrightarrow
 \nu_\Gamma\text{ satisfies }c.
\]
Therefore, the third rule gives
\[
 \mathsf{Bad}\in\PM(\PiP\cup\Dx\cup\Gamma)
 \Longleftrightarrow
 \nu_\Gamma\not\models\varphi.
\]
Since $\mathsf{Switch}\notin E\toggle\Gamma$, the goal is false before the
candidate toggle.  After toggling $\tau$, the final rule gives
\[
 \Ans(E\toggle(\Gamma\cup\{\tau\}))=1
 \Longleftrightarrow
 \nu_\Gamma\models\varphi.
 \tag{61}\label{eq:sat-cause}
\]
If $\varphi$ is satisfiable, the true-variable atoms of a satisfying
assignment form a witnessing contingency.  Conversely, any witnessing
contingency excludes $\tau$ and determines, by
Equation~\eqref{eq:sat-cause}, a satisfying assignment.  Thus, $\tau$ is a
cause exactly when $\varphi$ is satisfiable.  The construction has one
mutable assignment atom per variable, one switch atom, and one exogenous fact
per variable, clause, and literal occurrence, and therefore has polynomial
size.

\emph{Item 2: hardness of $\Robust$.}
We reduce from $\VC$ restricted to nonempty graphs.  The restriction remains
NP-complete under the map
$(G,k)\mapsto(G\mathbin{\dot\cup}K_2,k+1)$: a cover of $G$ extends by one
endpoint of the new edge, and every cover of the disjoint union contains an
endpoint of that edge whose removal leaves a cover of $G$.

Let $G=(V,E_G)$ be a nonempty graph.  Choose one orientation of each
undirected edge and include the corresponding facts $\mathsf{Edge}(u,v)$ in
$\Dx$.  Let
\[
 \Univ=E=\{\mathsf{Keep}(v)\mid v\in V\} \quad\text{and}\quad A=\Goal,
\]
and use exactly the fixed program in Equation~\eqref{eq:cover-program}.  It is
safe, nonrecursive, and stratified with $\mathsf{Bad}$ below $\Goal$.  Since
$G$ is nonempty and every keep-atom is initially present, the observed goal
is false.

For an intervention $J\subseteq\Univ$, let
$S_J=\{v\mid\mathsf{Keep}(v)\in J\}$.  Every atom in $J$ is initially
present, thus applying $J$ deletes exactly the keep-atoms of $S_J$.  The two
rules of~\eqref{eq:cover-program} yield
\[
\begin{aligned}
 \Ans(E\toggle J)=1
 &\Longleftrightarrow
 \neg\exists\{u,v\}\in E_G\;(u\notin S_J\land v\notin S_J)\\
 &\Longleftrightarrow
 \forall\{u,v\}\in E_G\;(u\in S_J\lor v\in S_J)\\
 &\Longleftrightarrow
 S_J\text{ is a vertex cover of }G.
\end{aligned}
 \tag{62}\label{eq:robust-vc}
\]
A cover of size at most $k$ therefore gives an outcome-changing intervention
of size at most $k$, and every such intervention yields a cover of the same
size.  Hence, $\Rob(E)\leq k$ exactly when $(G,k)$ is a yes-instance of
$\VC$.  The construction contains $|V|$ mutable atoms and $|E_G|$
exogenous edge facts, and is linear.

\emph{Item 3: hardness of $\Responsibility$.}
Use the same oriented-edge database and the first rule of
Equation~\eqref{eq:cover-program}, and replace its goal rule by the main-text
rule
\[
 \Goal\leftarrow\mathsf{Switch},\neg\mathsf{Bad}.
 \tag{63}\label{eq:responsibility-vc}
\]
Let
\[
 \Univ=\{\mathsf{Keep}(v)\mid v\in V\}\cup\{\mathsf{Switch}\},
 \quad
 E=\{\mathsf{Keep}(v)\mid v\in V\},
 \quad
 A=\Goal
 \quad\text{and}\quad
 \tau=\mathsf{Switch}.
\]
The switch is absent, so the observed goal is false.  A contingency
$\Gamma\subseteq\Univ\setminus\{\tau\}$ deletes precisely the keep-atoms
of
$S_\Gamma=\{v\mid\mathsf{Keep}(v)\in\Gamma\}$, and the goal remains false
before the candidate toggle.  After toggling $\tau$,
\[
\begin{aligned}
 \Ans(E\toggle(\Gamma\cup\{\tau\}))=1
 &\Longleftrightarrow
 \mathsf{Bad}\notin\PM(\PiP\cup\Dx\cup(E\toggle\Gamma))\\
 &\Longleftrightarrow
 S_\Gamma\text{ is a vertex cover of }G,
\end{aligned}
 \tag{64}\label{eq:responsibility-cover}
\]
where the second equivalence is Equation~\eqref{eq:robust-vc}.  A cover of
size at most $k$ yields a contingency of size at most $k$, and every
witnessing contingency yields a cover of the same size.  Thus,
$\Kappa_E(\tau)$ equals the minimum vertex-cover size.  The construction is
linear in $|V|+|E_G|$, and the program and goal predicate are fixed.

The NP upper bounds and the three polynomial reductions prove the theorem.
}
\subsubsection*{Proof of Theorem~\ref{thm:response-equivalence}}
Both target programs are fixed, safe, nonrecursive, and stratified.

For membership, the complement guesses $F\subseteq\Univ$.  Fixed-program
perfect-model evaluation computes $\Ans_{I_1}(F)$ and
$\Ans_{I_2}(F)$ in polynomial time.  The verifier accepts exactly when the
values differ.  Hence, nonequivalence is in NP, and
$\ResponseEquivalence$ is in coNP.

For hardness, let $\varphi$ be a CNF formula with variables
$x_1,\ldots,x_n$ and clauses $c_1,\ldots,c_m$.  Let $\Dx$ contain the facts
$\mathsf{Var}(x_i)$, $\mathsf{Clause}(c_j)$,
$\mathsf{Pos}(c_j,x_i)$ for positive occurrences, and
$\mathsf{Neg}(c_j,x_i)$ for negative occurrences.  Let
\[
 \Univ=\{\mathsf{True}(x_i){\mid}1\leq i\leq n\}.
\]
For every reduction instance, let $A=\Goal$.  The first fixed program is
\[
\begin{array}{rcl}
 \mathsf{Sat}(c)&\leftarrow&\mathsf{Pos}(c,x),\mathsf{True}(x),\\
 \mathsf{Sat}(c)&\leftarrow&\mathsf{Neg}(c,x),\mathsf{Var}(x),
                              \neg\mathsf{True}(x),\\
 \mathsf{Bad}&\leftarrow&\mathsf{Clause}(c),\neg\mathsf{Sat}(c),\\
 \Goal&\leftarrow&\neg\mathsf{Bad}.
\end{array}
\tag{58}\label{eq:equivalence-reduction}
\]
The final rule is ground and therefore safe.  A stratification places extensional predicates in
stratum $0$,
$\mathsf{Sat}$ in stratum $1$, $\mathsf{Bad}$ in stratum $2$, and $\Goal$ in stratum $3$.  The program is nonrecursive.

For $F\subseteq\Univ$, let assignment $\nu_F$ make $x_i$ true exactly
when $\mathsf{True}(x_i)\in F$.  For each clause $c_j$, the first rule
derives $\mathsf{Sat}(c_j)$ exactly when $c_j$ has a true positive
{literal}, and the second rule derives it exactly when $c_j$ has a true
{negative literal}.  Therefore,
\[
 \mathsf{Sat}(c_j)\in\PM(\PiP_1\cup\Dx\cup F)
 \Longleftrightarrow
 \nu_F\text{ satisfies }c_j.
\]
The third rule derives $\mathsf{Bad}$ exactly when some clause is not
satisfied.  The final rule gives
\[
 \Ans_{I_1}(F)=1
 \Longleftrightarrow
 \nu_F\models\varphi.
 \tag{59}\label{eq:equivalence-sat}
\]
{Let $\PiP_2=\emptyset$ over the same common extensional signature and
with the same designated goal atom $A=\Goal$.}  Since $\PiP_2$ has no rule
with head $\Goal$, $\Ans_{I_2}(F)=0$ for every $F\subseteq\Univ$.  If
$\varphi$ is unsatisfiable, Equation~\eqref{eq:equivalence-sat} gives
$\Ans_{I_1}(F)=\Ans_{I_2}(F)=0$ for every $F$.  If the responses are
equal, then $\Ans_{I_1}(F)=0$ for every assignment state $F$, and
Equation~\eqref{eq:equivalence-sat} implies that $\varphi$ is
unsatisfiable.  The two implications prove both correctness directions.

The construction adds one {mutable atom} per variable and one exogenous fact per
variable, clause, and {literal occurrence}.  Its size is linear in the input
CNF encoding.  Therefore, the reduction is polynomial, and
$\ResponseEquivalence$ is coNP-hard.

\vspace{0.25cm}
\section{Proofs of Supplementary Results}
\vspace{0.25cm}

\subsubsection*{Proof of Proposition~\ref{prop:separation}}
{
Let $\Univ=\{p,q,r\}$, where $p=P(c)$, $q=Q(c)$, and
$r=R(c)$.  The two rules of~\eqref{eq:separation-program} are ground,
nonrecursive, and safe.  We place the extensional predicates in stratum $0$
and $\Goal$ in stratum $1$.

For every $F\subseteq\Univ$, the first rule derives $\Goal$ exactly when
$p\in F$, $q\notin F$, and $r\notin F$.  The second rule derives
$\Goal$ exactly when $p,q,r\in F$.  Hence,
\[
 \Ans(F)=1
 \Longleftrightarrow
 F=\{p\}\ \text{or}\ F=\{p,q,r\}.
 \tag{30}\label{eq:sep-response}
\]

We first refute~\eqref{eq:support-decomposition}.  The state $\{p\}$ has
outcome $1$, whereas its only proper substate $\emptyset$ has outcome
$0$.  Thus, $\{p\}$ is a minimal positive support.  The state
$\{p,q\}$ contains that support but has outcome $0$ by
\eqref{eq:sep-response}.  Consequently, support containment does not imply
outcome $1$, and~\eqref{eq:support-decomposition} fails.

Let the observed state be $E=\emptyset$.  Since $E\toggle J=J$,
Equation~\eqref{eq:sep-response} gives
\[
 \Ans(E\toggle J)=1
 \Longleftrightarrow
 J=\{p\}\ \text{or}\ J=\{p,q,r\}.
\]
The intervention $\{p\}$ is inclusion-minimal, while
$\{p,q,r\}$ is not inclusion-minimal because
$\{p\}\subsetneq\{p,q,r\}$ already changes the outcome.  Therefore,
$\Flip_I(E)=\{\{p\}\}$.
In particular, $r$ belongs to no inclusion-minimal outcome-changing
intervention.

Let $\Gamma=\{p,q\}$.  We have $r\notin\Gamma$ and
\[
 \Ans(E\toggle\Gamma)=\Ans(\{p,q\})=0
  \quad\text{and}\quad
 \Ans(E\toggle(\Gamma\cup\{r\}))=\Ans(\{p,q,r\})=1.
\]
Thus, $\Gamma$ satisfies~\eqref{eq:cause}, and $r$ is an actual cause.
The same instance establishes both claims of the proposition.
}

\subsubsection*{Proof of Proposition~\ref{prop:sign-necessity}}
{
Let $p=P(c)$, $q=Q(c)$, and
$\PiP=\{\Goal\leftarrow P(c),\neg Q(c)\}$.  The program is ground,
safe, nonrecursive, and stratified.  Direct perfect-model evaluation gives
\[
\begin{array}{|c|c|}
\hline
 F & \Ans(F)\\
\hline
 \emptyset & 0\\
\hline
 \{p\} & 1\\
\hline
 \{q\} & 0\\
\hline
 \{p,q\} & 0\\
\hline
\end{array}
\tag{31a}\label{eq:sign-table}
\]
Let $C=(\{p\},\{q\})$.  Every state satisfying $C$ equals $\{p\}$,
thus $C$ is a $1$-implicant.  Removing the positive literal $p$ yields
$(\emptyset,\{q\})$, which is satisfied by $\emptyset$ of outcome $0$.
Removing the negative literal $\neg q$ yields
$(\{p\},\emptyset)$, which is satisfied by $\{p,q\}$ of outcome $0$.
Thus, both literals are necessary and $C$ is prime.

Every $1$-implicant must contain the positive literal $p$: otherwise, it
has a satisfying state omitting $p$, and both such states in
Table~\ref{eq:sign-table} have outcome $0$.  Every $1$-implicant must also
contain the negative literal $\neg q$: otherwise, it has a satisfying state
containing both $p$ and $q$, whose outcome is $0$.  Thus,
\[
 \Prime_I^1=\{(\{p\},\{q\})\}.
\]

For outcome $0$, let
\[
 D_p=(\emptyset,\{p\})  \quad\text{and}\quad
 D_q=(\{q\},\emptyset).
\]
Every state satisfying $D_p$ omits $p$ and has outcome $0$.  Removing
$\neg p$ yields the empty term, which admits $\{p\}$ of outcome $1$.
Thus, $D_p$ is prime.  Every state satisfying $D_q$ contains $q$ and
has outcome $0$.  Removing $q$ again yields the empty term and admits
$\{p\}$.  Thus, $D_q$ is prime.

Let $D$ be any $0$-implicant.  If $D$ contains neither $\neg p$ nor
$q$, then $\{p\}$ satisfies $D$, contradicting its outcome $1$.
Thus, $D_p\preceq D$ or $D_q\preceq D$.  Primeness permits no proper
implicant subterm, thus every prime $0$-implicant equals $D_p$ or $D_q$.
Therefore,
\[
 \Prime_I^0=\{D_p,D_q\}.
\]
The unique prime truth implicant contains the negative literal $\neg q$, and
the prime falsity implicant $D_q$ contains the positive literal $q$.
}

\subsubsection*{Proof of Proposition~\ref{prop:calculus}}
{
We prove the two product identities by both inclusions and verify primeness in
each direction.  The remaining identities follow from duality and Boolean
negation.

\emph{Conjunction truth implicants.}
Let
\[
 \mathcal M_{11}=
 \{C_f\merge C_g\mid C_f\in\Prime^1(f),\ C_g\in\Prime^1(g),
                    \ C_f,C_g\text{ compatible}\}.
\]
Every $M\in\mathcal M_{11}$ is a $1$-implicant of $f\land g$: a state
satisfying $M$ satisfies both constituent terms, thus it has
$f=g=1$.

Let $C\in\Prime^1(f\land g)$.  Since $C$ forces $f\land g=1$, it
separately forces $f=1$ and $g=1$.  Finite minimization among the subterms
of $C$ yields $C_f\in\Prime^1(f)$ and $C_g\in\Prime^1(g)$ with
$C_f\preceq C$ and $C_g\preceq C$.  Both are subterms of the common term
$C$, thus they are compatible and
$M_0=C_f\merge C_g\preceq C$.  Choose a $\preceq$-minimal member $M$
of $\mathcal M_{11}$ satisfying $M\preceq M_0$; such a member exists
because $M_0$ belongs to the finite family.  No member of
$\mathcal M_{11}$ is properly below $M$, since any such member would also
be below $M_0$.  Thus, $M\in\Min_{\preceq}(\mathcal M_{11})$.  The first
paragraph shows that $M$ is a $1$-implicant of $f\land g$.  Since
$M\preceq C$ and $C$ is prime, $M=C$.  Hence,
\[
 \Prime^1(f\land g)
 \subseteq
 \Prime^1(f)\otimes\Prime^1(g).
\]

For the reverse inclusion, let
$M\in\Prime^1(f)\otimes\Prime^1(g)$.  The first paragraph shows that $M$
is a $1$-implicant of $f\land g$.  Suppose a proper subterm
$M'\prec M$ were also a $1$-implicant.  The term $M'$ separately forces $f=1$ and $g=1$.  Among its
finitely many subterms that force $f=1$, choose a prime one
$C_f'\in\Prime^1(f)$; among its subterms that force $g=1$, choose a
prime one $C_g'\in\Prime^1(g)$.  Then,
$C_f',C_g'\preceq M'$.  Their merge $M_0'$ belongs to
$\mathcal M_{11}$ and satisfies $M_0'\preceq M'\prec M$.  A minimal
member of $\mathcal M_{11}$ below $M_0'$ belongs to the product and is
properly below $M$, contradicting the defining minimality of $M$.
Therefore, $M$ is prime, and
\[
 \Prime^1(f\land g)=\Prime^1(f)\otimes\Prime^1(g).
\]

\emph{Disjunction falsity implicants.}
Let
\[
 \mathcal M_{00}=
 \{D_f\merge D_g\mid D_f\in\Prime^0(f),\ D_g\in\Prime^0(g),
                    \ D_f,D_g\text{ compatible}\}.
\]
Every member of $\mathcal M_{00}$ forces $f=0$ and $g=0$, and therefore
forces $f\lor g=0$.

Let $D\in\Prime^0(f\lor g)$.  The term $D$ separately forces $f=0$
and $g=0$.  Choose prime subterms $D_f\in\Prime^0(f)$ and
$D_g\in\Prime^0(g)$ below $D$.  They are compatible, their merge is below
$D$, and a minimal member of $\mathcal M_{00}$ below that merge is a
$0$-implicant of $f\lor g$.  Primeness of $D$ forces equality, yielding
\[
 \Prime^0(f\lor g)
 \subseteq
 \Prime^0(f)\otimes\Prime^0(g).
\]

Conversely, let $M\in\Prime^0(f)\otimes\Prime^0(g)$.  It is a
$0$-implicant of $f\lor g$.  If a proper subterm $M'\prec M$ were a
$0$-implicant, then $M'$ would separately force $f=0$ and $g=0$.
Prime minimization below $M'$ would produce a member of
$\mathcal M_{00}$, and then a product member, properly below $M$, a
contradiction.  Hence, $M$ is prime, and
\[
 \Prime^0(f\lor g)=\Prime^0(f)\otimes\Prime^0(g).
\]

\emph{Remaining identities.}
Proposition~\ref{prop:representation} gives
$\Prime^0(h)=\Dual(\Prime^1(h))$ and
$\Prime^1(h)=\Dual(\Prime^0(h))$ for every Boolean response $h$.  Applying
these equalities to the two identities already proved yields
\[
\begin{aligned}
 \Prime^0(f\land g)
 &=\Dual\bigl(\Prime^1(f\land g)\bigr)
  =\Dual\bigl(\Prime^1(f)\otimes\Prime^1(g)\bigr),\\
 \Prime^1(f\lor g)
 &=\Dual\bigl(\Prime^0(f\lor g)\bigr)
  =\Dual\bigl(\Prime^0(f)\otimes\Prime^0(g)\bigr).
\end{aligned}
\]
Finally, for every term $C$,
\[
\begin{aligned}
 C\text{ is a }b\text{-implicant of }\neg f
 &\Longleftrightarrow
 \forall F\models C,\ 1-f(F)=b\\
 &\Longleftrightarrow
 \forall F\models C,\ f(F)=1-b\\
 &\Longleftrightarrow
 C\text{ is a }(1-b)\text{-implicant of }f.
\end{aligned}
\]
The equivalence is unchanged for every proper subterm of $C$, so it also
preserves primeness.  Therefore,
$\Prime^b(\neg f)=\Prime^{1-b}(f)$.
}

\subsubsection*{Proof of Proposition~\ref{prop:relevance}}
{
Let Conditions 1--4 be those in Proposition~\ref{prop:relevance}.
Condition 1 and Condition 4 are identical after taking the empty contingency
in Definition~\ref{def:cause}.  We prove
$1\Rightarrow2$, $1\Rightarrow3$, $2\Rightarrow1$, and
$3\Rightarrow1$.

Assume Condition 1.  Choose $F\subseteq\Univ$ with
$\Ans(F)\neq\Ans(F\toggle\{\tau\})$, and let $b=\Ans(F)$.  By the prime
cover, choose $C\in\Prime_I^b$ with $F\models C$.  If $C$ contained no
literal on $\tau$, toggling $\tau$ would preserve satisfaction of every
literal of $C$, thus $F\toggle\{\tau\}\models C$.  The implicant property
would then give $\Ans(F\toggle\{\tau\})=b$, a contradiction.  Thus, $C$
contains a literal on $\tau$.  If $b=1$, then
$C\in\Prime_I^1$ establishes Condition 2.  If $b=0$, then
$C\in\Prime_I^0$ establishes Condition 3.

Let $G=F\toggle\{\tau\}$.  Its outcome is $1-b$.  By the prime cover,
choose $D\in\Prime_I^{1-b}$ with $G\models D$.  If $D$ contained no
literal on $\tau$, then $F\models D$, because $F$ and $G$ differ only on
$\tau$.  The implicant property would give $\Ans(F)=1-b$, contradicting
$\Ans(F)=b$.  Thus, $D$ contains a literal on $\tau$.  If $b=1$, then
$D\in\Prime_I^0$ establishes Condition 3; if $b=0$, then
$D\in\Prime_I^1$ establishes Condition 2.  Therefore, Condition 1 implies
both Conditions 2 and 3 for both values of $b$.

Assume Condition 2.  Let $C\in\Prime_I^1$ contain a literal $L_\tau$ on
$\tau$, and let $C_{-\tau}$ be the proper subterm obtained by deleting
that literal.  Since $C$ is prime, $C_{-\tau}$ is not a $1$-implicant.
Thus, some state $G$ satisfies $C_{-\tau}$ and has outcome $0$.  If
$G$ satisfied $L_\tau$, then $G\models C$, contradicting the implicant
property of $C$.  Hence, $G$ satisfies the complementary literal on
$\tau$, and $G\toggle\{\tau\}\models C$.  We obtain
\[ \Ans(G)=0 \quad\text{and}\quad \Ans(G\toggle\{\tau\})=1,
\]
which is Condition 1.

Assume Condition 3.  Let $D\in\Prime_I^0$ contain a literal $L_\tau$, and
let $D_{-\tau}$ delete it.  Primeness gives a state $G\models D_{-\tau}$
with outcome $1$.  The state $G$ cannot satisfy $L_\tau$, because then
$G\models D$ would force outcome $0$.  Therefore,
$G\toggle\{\tau\}\models D$, and
\[
 \Ans(G)=1
  \quad\text{and}\quad
 \Ans(G\toggle\{\tau\})=0.
\]
Condition 1 follows.  All four conditions are equivalent.

{We finally prove the quantified consequence.  Let
$F,J\subseteq\Univ$, and enumerate
$J\setminus\Rel_I=\{u_1,\ldots,u_m\}$.  Let
\[
 J_0=J,  \quad\text{and}\quad
 J_j=J_{j-1}\setminus\{u_j\}
 \quad(1\leq j\leq m).
\]
For $0\leq j\leq m$, let $G_j=F\toggle J_j$.  The states
$G_{j-1}$ and $G_j$ differ only on $u_j$.  Since
$u_j\notin\Rel_I$, Condition 1 fails at every state, and therefore
$\Ans(G_{j-1})=\Ans(G_j)$.  Induction on $j$ yields
\[
 \Ans(F\toggle J)=\Ans(F\toggle J_m).
\]
Since $J_m=J\cap\Rel_I$, we obtain
$\Ans(F\toggle J)=\Ans(F\toggle(J\cap\Rel_I))$.}
}

\subsubsection*{Proof of Corollary~\ref{cor:invariance}}
{
Let $I_1$ and $I_2$ share $\Dx$, $\Univ$, $E$, and $A$.

Assume $\Prime_{I_1}^1=\Prime_{I_2}^1$.  For every state
$F\subseteq\Univ$, the prime cover gives
\[
\begin{aligned}
 \Ans_{I_1}(F)=1
 &\Longleftrightarrow
 \exists C\in\Prime_{I_1}^1\;(F\models C)\\
 &\Longleftrightarrow
 \exists C\in\Prime_{I_2}^1\;(F\models C)\\
 &\Longleftrightarrow
 \Ans_{I_2}(F)=1.
\end{aligned}
\]
The responses are Boolean, so they agree on $F$.  Since $F$ was arbitrary,
they agree on every mutable state.  If $\Prime_{I_1}^0=\Prime_{I_2}^0$, then for every
$F\subseteq\Univ$,
\[
\begin{aligned}
 \Ans_{I_1}(F)=0
 &\Longleftrightarrow
 \exists C\in\Prime_{I_1}^0\;(F\models C)\\
 &\Longleftrightarrow
 \exists C\in\Prime_{I_2}^0\;(F\models C)\\
 &\Longleftrightarrow
 \Ans_{I_2}(F)=0.
\end{aligned}
\]
Booleanity again gives response equality.

Conversely, assume
$\Ans_{I_1}(F)=\Ans_{I_2}(F)$ for every $F\subseteq\Univ$.  For every
term $C$ and $b\in\{0,1\}$,
\[
\begin{aligned}
 C\text{ is a }b\text{-implicant of }I_1
 &\Longleftrightarrow
 \forall F\models C,\ \Ans_{I_1}(F)=b\\
 &\Longleftrightarrow
 \forall F\models C,\ \Ans_{I_2}(F)=b\\
 &\Longleftrightarrow
 C\text{ is a }b\text{-implicant of }I_2.
\end{aligned}
\]
The two instances use the same subterm order.  Thus, a $b$-implicant has a
proper $b$-implicant subterm in one instance exactly when it has one in the
other.  Thus,
$\Prime_{I_1}^b=\Prime_{I_2}^b$ for $b=0,1$.  Equation~\eqref{eq:prime-duality}
also proves directly that equality of either polarity is equivalent to equality
of the other.

Assume the equivalent response-equality conditions.  Let $\tau\in\Univ$ and
$\Gamma\subseteq\Univ\setminus\{\tau\}$.  Both instances assign the same
outcome to
$E\toggle\Gamma$ and to
$E\toggle(\Gamma\cup\{\tau\})$.  Therefore, $\Gamma$ satisfies the two
causality equations for $I_1$ exactly when it satisfies them for $I_2$.
The feasible contingency families are identical for every candidate.  Their
minimum cardinalities are consequently equal, including the common value
$\Inf$ when the family is empty.  Equation~\eqref{eq:responsibility} then
gives equal responsibilities and equal actual-cause families.

Likewise, for every intervention $J\subseteq\Univ$,
\[
 \Ans_{I_1}(E\toggle J)\neq\Ans_{I_1}(E)
 \,\Longleftrightarrow\,
 \Ans_{I_2}(E\toggle J)\neq\Ans_{I_2}(E).
\]
Thus, the outcome-changing intervention families coincide, and their minimum
cardinalities, including $\Inf$, are equal.  The robustness radii agree.
}

\subsubsection*{Proof of Corollary~\ref{cor:positive-polarity}}
{
Let $b=\Ans(E)$.

Assume first that $b=1$.  Let $\tau\notin E$, and let
$\Gamma\subseteq\Univ\setminus\{\tau\}$ be arbitrary.  Since
$\tau\notin E$ and $\tau\notin\Gamma$, the state
$F=E\toggle\Gamma$ omits $\tau$, and
$F\toggle\{\tau\}=F\cup\{\tau\}$.  If $\Gamma$ preserved the observed
outcome, then $\Ans(F)=1$.  Monotonicity gives
\[
 1=\Ans(F)\leq\Ans(F\cup\{\tau\})\leq1,
\]
so the candidate toggle cannot produce outcome $0$.  Thus, no atom outside
$E$ is a cause, and every actual cause belongs to $E$.

Let $\tau\in E$.  Then, $\ell_E(\tau)=\tau$.  By
Theorem~\ref{thm:causal-characterization}, $\tau$ is a cause exactly when a
prime $1$-implicant contains the positive literal $\tau$.  By
Corollary~\ref{cor:positive}, the prime $1$-implicants are exactly
$(S,\emptyset)$ for $S\in\Supp_I$.  Therefore,
\[
 \tau\text{ is a cause}
 \Longleftrightarrow
 \exists S\in\Supp_I\;(\tau\in S).
\]

Assume now that $b=0$.  Let $\tau\in E$, and let
$\Gamma\subseteq\Univ\setminus\{\tau\}$.  The state
$F=E\toggle\Gamma$ contains $\tau$, and
$F\toggle\{\tau\}=F\setminus\{\tau\}$.  If $\Gamma$ preserved the
observed outcome, then $\Ans(F)=0$.  Monotonicity gives
\[
 0\leq\Ans(F\setminus\{\tau\})\leq\Ans(F)=0,
\]
thus deleting $\tau$ cannot produce outcome $1$.  Thus, every actual cause
lies in $\Univ\setminus E$.

Let $\tau\notin E$.  Then, $\ell_E(\tau)=\neg\tau$.  By
Theorem~\ref{thm:causal-characterization}, $\tau$ is a cause exactly when a
prime $0$-implicant contains $\neg\tau$.  Corollary~\ref{cor:positive}
identifies every prime $0$-implicant with
$(\emptyset,T)$ for $T\in\Trans(\Supp_I)$.  Such a term contains
$\neg\tau$ exactly when $\tau\in T$.  Consequently,
\[
 \tau\text{ is a cause}
 \Longleftrightarrow
 \exists T\in\Trans(\Supp_I)\;(\tau\in T).
\]
The two outcome cases and both candidate polarities are exhausted.
}

\subsubsection*{Proof of Proposition~\ref{prop:circuit}}
{
Let $\Delta$ be the active domain of $(\PiP,\Dx,\Univ,A)$, and let
$N=|\Delta|$.  Since $\PiP$ is fixed, predicate arities, rule lengths, and
the numbers of variables per rule are constants.  If a rule has at most $d$
variables, it has at most $N^d$ ground instances.  Thus, the complete
grounding contains polynomially many ground atoms and ground rules.

For every ground extensional atom $L$, construct its gate as follows.  If
$L\in\Univ$, use the input gate $X_L$.  If $L\in\Dx$, use constant
$1$.  Every other extensional atom uses constant $0$.  Under input
$\chi_F$, these gates compute membership in $\Dx\cup F$.

Renumber the stratification so that all intensional predicates have positive
strata and the used intensional strata are $1,\ldots,s$.  We prove gate correctness by
outer induction on the stratum.  The outer induction hypothesis states that,
for every ground atom in a stratum below $i$, its final gate computes its
completed perfect-model truth value for every input state.  The hypothesis is
vacuous for the least intensional stratum.

Let $B_i$ be the family of ground intensional atoms in stratum $i$, and let
$m_i=|B_i|$.  For $H\in B_i$, let $G_{H,0}=0$.  For
$1\leq t\leq m_i$, let
\[
 G_{H,t}=G_{H,t-1}\vee
 \bigvee_{\rho\in\mathsf{Gr}_i(H)}
 \mathsf{Body}_{\rho,t-1},
 \tag{56}\label{eq:round-gate}
\]
where $\mathsf{Gr}_i(H)$ is the family of ground stratum-$i$ rules with
head $H$.  The empty disjunction is $0$, and an empty body has gate
$1$.

For a nonempty ground body, use a conjunction of one gate per body atom.  The
following cases exhaust safe stratified rule syntax.
\begin{enumerate}
\item A positive extensional atom uses its input or constant membership gate.
\item A positive lower-stratum atom uses its completed final gate.
\item A positive stratum-$i$ atom uses its round-$(t-1)$ gate.
\item A negated extensional atom uses the Boolean negation of its extensional
membership gate.
\item A negated intensional atom belongs to a lower stratum by stratification
and uses the Boolean negation of its completed final gate.
\end{enumerate}
No negated current- or higher-stratum atom is permitted by stratification.

For a state $F\subseteq\Univ$, let $T_{i,F}$ be the inflationary
immediate-consequence operator for stratum $i$, with extensional facts and
completed lower strata held constant.  We prove simultaneously for every
$H\in B_i$ and every $t\in\{0,\ldots,m_i\}$ that
\[
 G_{H,t}(\chi_F)=1
 \Longleftrightarrow
 H\in T_{i,F}^t(\emptyset).
 \tag{57}\label{eq:round-induction}
\]

For $t=0$, the left side is false because $G_{H,0}=0$, and the right side
is false because $T_{i,F}^0(\emptyset)=\emptyset$.

Assume~\eqref{eq:round-induction} for every stratum-$i$ atom at round
$t-1$.  Let $\rho$ be a ground rule.  We verify its body gate atom by
atom.  Case 1 is true exactly when the extensional atom belongs to
$\Dx\cup F$, by the extensional gate definition.  Case 2 is true exactly
when the lower-stratum atom belongs to its completed relation, by the outer
induction hypothesis.  Case 3 is true exactly when the atom belongs to
$T_{i,F}^{t-1}(\emptyset)$, by the inner induction hypothesis.  Cases 4 and
5 negate the exact truth values from Cases 1 and 2.  A conjunction gate is
therefore true exactly when every positive body atom is present and every
negated body atom is absent in the round-$(t-1)$ interpretation.  This is
exactly the satisfaction condition for the body of $\rho$.

Equation~\eqref{eq:round-gate} is true exactly when either $H$ was present at
round $t-1$, or some rule with head $H$ has a body true at that round.
This condition is equivalent to
$H\in T_{i,F}^t(\emptyset)$, proving the inner induction step.

The operator $T_{i,F}$ is inflationary on the finite family $B_i$.  Every
strict increase adds at least one new atom, so it has at most $m_i$ strict
increases.  Hence,
\[
 T_{i,F}^{m_i}(\emptyset)=T_{i,F}^{m_i+1}(\emptyset).
\]
The stabilized relation is the least fixpoint for stratum $i$.  Let
$G_H=G_{H,m_i}$.  Equation~\eqref{eq:round-induction} shows that $G_H$
computes the completed truth value of $H$.  Thus, the outer induction
hypothesis holds for stratum $i$.  Induction over all strata proves that the
final gate for $A$, denoted $\Circ_{\PiP,\Dx,\Univ,A}$, satisfies
\eqref{eq:circuit} for every state $F$.

For the size bound, let $R_i$ be the number of ground rules in stratum $i$,
and let $b_0$ be the maximum rule-body length, a constant.  At each of the
$m_i$ rounds, the construction uses at most $O(R_i b_0)$ body and
combination gates.  Thus, stratum $i$ uses $O(m_iR_i b_0)$ gates.  The
number of strata is constant, and every $m_i$ and $R_i$ is polynomial in
the grounding size.  The total circuit size and construction time are therefore
polynomial in $|\Dx|+|\Univ|+|A|$.
}

\subsubsection*{Proof of Proposition~\ref{prop:solver-encoding}}
{
Let an assignment to the circuit inputs represent
$ F_X=\{u\in\Univ\mid X_u=1\}$.
For $e_u=\chi_E(u)$, the Boolean value $X_u\oplus e_u$ equals $1$
exactly when $u$ belongs to the symmetric difference $F_X\toggle E$.
Therefore,
\[
 \sum_{u\in\Univ}(X_u\oplus e_u)=|F_X\toggle E|.
 \tag{60}\label{eq:hamming-identity}
\]
Proposition~\ref{prop:circuit} gives
$\Circ(X)=\Ans(F_X)$.

\emph{Item 1, forward direction.}
Assume~\eqref{eq:robust-encoding} is satisfiable, and let $X$ be a satisfying
assignment.  Let $F=F_X$ and $J=E\toggle F$.  Symmetric-difference
identities give $E\toggle J=F$.  Equation~\eqref{eq:hamming-identity} and the
cardinality constraint give $|J|\leq k$.  The output constraint gives
\[
 \Ans(E\toggle J)=\Ans(F)=\Circ(X)=1-b.
\]
Thus, $J$ is an outcome-changing intervention of size at most $k$, and
$\Rob(E)\leq k$.

\emph{Item 1, reverse direction.}
Assume $\Rob(E)\leq k$.  Choose $J\subseteq\Univ$ with
$|J|\leq k$ and $\Ans(E\toggle J)=1-b$.  Let
$F=E\toggle J$ and $X=\chi_F$.  Circuit correctness gives
$\Circ(X)=1-b$.  Since $F\toggle E=J$,
Equation~\eqref{eq:hamming-identity} gives
\[
 \sum_{u\in\Univ}(X_u\oplus e_u)=|J|\leq k.
\]
Thus, $X$ satisfies~\eqref{eq:robust-encoding}.

\emph{Item 2, forward direction.}
Assume~\eqref{eq:responsibility-encoding} is satisfiable, and let $X$ be a
satisfying assignment.  Let $F=F_X$ and $\Gamma=E\toggle F$.  The equality
$X_\tau=e_\tau$ means that $F$ and $E$ agree on $\tau$, thus
$\tau\notin\Gamma$.  The first output constraint gives
\[
 \Ans(E\toggle\Gamma)=\Ans(F)=b.
\]
Complementing input coordinate $\tau$ represents
\[
 F\toggle\{\tau\}
 =E\toggle(\Gamma\cup\{\tau\}),
\]
because $\tau\notin\Gamma$.  The second output constraint therefore gives
\[
 \Ans(E\toggle(\Gamma\cup\{\tau\}))=1-b.
\]
For every $u\neq\tau$, the summand $X_u\oplus e_u$ indicates membership
of $u$ in $\Gamma$, while the candidate summand is zero.  Hence, the
cardinality sum equals $|\Gamma|\leq k$.  Thus, $\Gamma$ is a witnessing
contingency and $\Kappa_E(\tau)\leq k$.

\emph{Item 2, reverse direction.}
Assume $\Kappa_E(\tau)\leq k$.  Choose a contingency
$\Gamma\subseteq\Univ\setminus\{\tau\}$ with
$|\Gamma|\leq k$ satisfying~\eqref{eq:cause}.  Let
$F=E\toggle\Gamma$ and $X=\chi_F$.  Since $\tau\notin\Gamma$,
$X_\tau=e_\tau$.  Circuit correctness and~\eqref{eq:cause} give
\[
 \Circ(X)=b \quad\text{and}\quad
 \Circ(X^{\oplus\tau})=1-b.
\]
Equation~\eqref{eq:hamming-identity}, restricted to
$\Univ\setminus\{\tau\}$, gives the cardinality value
$|\Gamma|\leq k$.  Hence, $X$ satisfies
\eqref{eq:responsibility-encoding}.

If the cardinality constraint is removed from
\eqref{eq:responsibility-encoding}, satisfiability is equivalent to the
existence of some contingency, with no bound on its size.  This is exactly the
decision problem $\Cause$.

For a CNF translation, introduce one Tseitin variable for every circuit gate and
clauses enforcing the corresponding gate equivalence.  Introduce deviation
variables
$Y_u\leftrightarrow(X_u\oplus e_u)$, and encode the relevant cardinality
constraint with any polynomial-size Boolean cardinality encoding.  These
clauses preserve satisfiability in both directions because every assignment of
the input variables extends uniquely to the gate values, and every satisfying
Tseitin assignment restricts to the encoded circuit computation.

For unweighted MaxSAT, keep all circuit, output, candidate, and coordinate
constraints hard, and add the unit soft clause $\neg Y_u$ for every permitted
deviation.  A feasible assignment violates exactly one soft clause per changed
atom.  {Under the convention that an infeasible hard part represents
value $\Inf$, the optimum represents $\Rob(E)$ for the robustness
encoding and $\Kappa_E(\tau)$ for the contingency encoding.}  The circuit and all auxiliary encodings have polynomial
size.
}

\subsubsection*{Proof of Proposition~\ref{prop:profile-separation}}
{
Let $\Univ=\{p,q,r\}$, $E=\{p\}$, and $\Dx=\emptyset$.  Let
$\PiP_1=\{A\leftarrow p\}$, and let
$\PiP_2=\PiP_1\cup\{A\leftarrow q,r\}$.  For every
$F\subseteq\Univ$,
\[
 \Ans_{I_1}(F)=1\Longleftrightarrow p\in F  \quad\text{and}\quad
 \Ans_{I_2}(F)=1\Longleftrightarrow
 p\in F\ \lor\ \{q,r\}\subseteq F.
 \tag{47}\label{eq:profile-responses}
\]
Both observed outcomes equal $1$.

For candidate $p$, the empty contingency satisfies
\[
 \Ans_{I_j}(E)=1  \quad\text{and}\quad
 \Ans_{I_j}(E\toggle\{p\})=\Ans_{I_j}(\emptyset)=0
 \quad\quad (j=1,2).
\]
Thus, $p$ is counterfactual in both instances,
\[
 \Kappa_E^{I_1}(p)=\Kappa_E^{I_2}(p)=0
 \quad\text{and}\quad
 \Resp_E^{I_1}(p)=\Resp_E^{I_2}(p)=1.
 \tag{48}\label{eq:profile-p}
\]

We prove that $q$ is not a cause in either instance.  Let
$\Gamma\subseteq\Univ\setminus\{q\}$, and let $F=E\toggle\Gamma$.  A
witnessing contingency would require $\Ans_{I_j}(F)=\Ans_{I_j}(E)=1$.
Since $q\notin F$, toggling $q$ produces $F\cup\{q\}$.  Both responses
in~\eqref{eq:profile-responses} are monotone, thus
\[
 \Ans_{I_j}(F)\leq\Ans_{I_j}(F\cup\{q\})
 \quad(j=1,2).
\]
The left side equals $1$, and thus the right side also equals $1$.
Toggling $q$ cannot reverse the observed outcome after any contingency.
Therefore, $q$ is not a cause in either instance.  For $r$, let $\Gamma\subseteq\Univ\setminus\{r\}$ and
$F=E\toggle\Gamma$.  A witnessing contingency requires $\Ans_{I_j}(F)=1$.
Since $r\notin F$, monotonicity gives
$\Ans_{I_j}(F)\leq\Ans_{I_j}(F\cup\{r\})$.  Thus, inserting $r$ also
preserves outcome $1$, and $r$ is not a cause in either instance.
Thus, both instances have exactly the cause $p$, and both assign
responsibility $0$ to $q$ and $r$.

Toggling $p$ changes the observed outcome in both instances, so each
robustness radius is at most $1$.  An intervention of size $0$ leaves the
observed true outcome unchanged, so each radius is at least $1$.  Hence,
\[
 \Rob_{I_1}(E)=\Rob_{I_2}(E)=1.
 \tag{49}\label{eq:profile-rob}
\]

At the state $F=\{q,r\}$, Equation~\eqref{eq:profile-responses} gives
$\Ans_{I_1}(F)=0$ and $\Ans_{I_2}(F)=1$.  Therefore, the two full
responses differ although their actual causes, responsibility values, and
robustness radii agree at $E$.
}

\subsubsection*{Proof of Theorem~\ref{thm:implicant-complexity}}
The input is $(\Dx,\Univ,A,C,b)$, while the safe nonrecursive stratified
program and the goal predicate are fixed.  We prove coNP membership by a
counterexample state and coNP-hardness by a reduction from CNF
unsatisfiability.

For membership, the complement guesses $F\subseteq\Univ$, verifies
$F\models C$, evaluates the fixed program on $\Dx\cup F$, and accepts
when $\Ans(F)=1-b$.  The witness has $|\Univ|$ bits, term satisfaction is
polynomial, and fixed-program evaluation has polynomial data complexity.
Thus, non-implicanthood is in NP, and $\Implicant$ is in coNP.

For hardness, let $\varphi$ be a CNF formula with variable family $V$ and
clause family $K$.  Let $\Dx$ contain the facts
$\mathsf{Var}(v)$, $\mathsf{Clause}(k)$,
$\mathsf{Pos}(k,v)$, and $\mathsf{Neg}(k,v)$ determined by the formula.
Let
$\Univ=\{\mathsf{True}(v){\mid}v\in V\}$, let $C=\epsilon$, let
$b=0$, and {let $A=\Goal$}.  Use the fixed program
\[
\begin{array}{rcl}
 \mathsf{Sat}(k)&\leftarrow&\mathsf{Pos}(k,v),\mathsf{True}(v),\\
 \mathsf{Sat}(k)&\leftarrow&\mathsf{Neg}(k,v),\mathsf{Var}(v),
                              \neg\mathsf{True}(v),\\
 \mathsf{Bad}&\leftarrow&\mathsf{Clause}(k),\neg\mathsf{Sat}(k),\\
 \Goal&\leftarrow&\neg\mathsf{Bad}.
\end{array}
\]
{The program is safe and nonrecursive.  A stratification places the
extensional predicates in stratum $0$, $\mathsf{Sat}$ in stratum $1$,
$\mathsf{Bad}$ in stratum $2$, and $\Goal$ in stratum $3$.}
For every state $F\subseteq\Univ$, the derivation used in
Equation~\eqref{eq:equivalence-sat} gives
\[
 \Ans(F)=1\Longleftrightarrow F\text{ encodes a satisfying assignment of }
 \varphi.
\]
Since every state satisfies the empty term,
\[\epsilon\text{ is a }0\text{-implicant}
 \,\,\Longleftrightarrow\,\,
 \Ans(F)=0\text{ for every }F\subseteq\Univ
 \,\,\Longleftrightarrow\,\,
 \varphi\text{ is unsatisfiable}.\]
The empty term has no proper subterm, so it is prime whenever it is an
implicant.  The construction contains one {mutable atom} per variable and one
exogenous fact per variable, clause, and {literal occurrence}.  It is
polynomial.  CNF unsatisfiability is coNP-complete, thus $\Implicant$ is
coNP-hard.

\end{document}